\documentclass[12pt]{article}
\usepackage{enumitem}
\usepackage[english]{babel}
\usepackage[utf8]{inputenc}
\usepackage[authoryear,comma]{natbib}
\usepackage{amsmath,multibib}
\usepackage{bbm}
 \usepackage{relsize}
\usepackage{amssymb,amsthm,mathabx,mathtools}
\usepackage{mathrsfs}
\usepackage{graphicx}
\usepackage{url}
\usepackage{caption,subcaption}
\usepackage{booktabs}
\usepackage{multirow}
\usepackage{caption}
\usepackage{epsfig,float}
\usepackage{url} % not crucial - just used below for the URL
\usepackage[usenames,dvipsnames]{color} 
\usepackage{listings}
\usepackage{mathrsfs,bm}

\usepackage{graphicx}

\theoremstyle{plain}
\newtheorem{thm}{Theorem}

\newcommand{\bthm}{\begin{thm}}
\newcommand{\ethm}{\end{thm}}
\newcommand{\bpf}{\begin{proof}}
\newcommand{\epf}{\end{proof}}
\theoremstyle{definition}
\newtheorem{defn}{Definition}
\newtheorem{example}{Example}
\newtheorem{rem}{Remark}
\usepackage{deep-macro}
\usepackage{epigraph}

\usepackage{mathtools}

\usepackage{epigraph}
\usepackage{framed}
\usepackage{relsize}
\usepackage[most]{tcolorbox}
\usepackage{tikz}
\usetikzlibrary{automata, positioning}
\usetikzlibrary{arrows, arrows.meta, calc, positioning, quotes, shapes, shapes.geometric}
\tikzstyle{block} = [rectangle, draw, text centered,    text width=7em, rounded corners, minimum height=1cm]
\tikzstyle{line} = [draw, -latex',line width=.22mm]
\usepackage{bbm}
\usepackage{tabularx}
\usepackage{tabularx}
\newcolumntype{Y}{>{\centering\arraybackslash}X}
\newcolumntype{f}{>{\centering\arraybackslash}X}
\newcolumntype{h}{>{\hsize=.5\hsize\centering\arraybackslash\extracolsep{.1em}}X}

\newcolumntype{C}[1]{>{\hsize=#1\hsize\centering\arraybackslash}X}

\usepackage{deep-macro}

\usepackage[bottom]{footmisc}
\usepackage{cancel}

\usepackage{JASA_manu}
\usepackage{setspace}

\newcommand\blfootnote[1]{%
  \begingroup
  \renewcommand\thefootnote{}\footnote{#1}%
  \addtocounter{footnote}{-1}%
  \endgroup
}

\begin{document}
\title{
\vspace{-2em}
Modelplasticity and Abductive Decision Making$^*$\blfootnote{The author owes a debt of gratitude to Prof. Stephen Stigler for  inspiring him to think about the problem.}
}
\author{
\centering
\begin{tabularx}{.97\linewidth}{ff}
Subhadeep (DEEP) Mukhopadhyay\\
\texttt{deep@unitedstatalgo.com} 
\end{tabularx} }

\date{}
\maketitle

%%-----------------------------------

\vspace{-2em}
\begin{abstract} 
`All models are wrong but some are useful' (George \citealt{box1979robustness}). But, how to find those useful ones starting from an imperfect model? How to make informed data-driven decisions equipped with an imperfect model? These fundamental questions appear to be pervasive in virtually all empirical fields---including economics, finance, marketing, healthcare, climate change, defense planning, and operations research. This article presents a modern approach (builds on two core ideas: abductive thinking and density-sharpening principle) and practical guidelines to tackle these issues in a systematic manner. 
\end{abstract}

\noindent\textsc{\textbf{Keywords}}:  Abductive Decision-making; Model Risk Management; The Uncertainty Principle; Density-Sharpening Principle; Creation of New Knowledge; Quantile Decision Analysis.

\vskip1em

% % \renewcommand{\baselinestretch}{0.25ex}
% % \setlength{\parskip}{0.1ex}
% %\renewcommand\contentsname{}

% \setcounter{tocdepth}{3}
% \begin{spacing}{0.74}
% \tableofcontents
% \end{spacing}
% %\pagenumbering{gobble}

% %\renewcommand{\baselinestretch}{1.6}
% \setlength{\parskip}{1.24ex}
% \setstretch{1.3}

%\renewcommand{\baselinestretch}{1.2}

%%%%%%%%%%%%%%%%%%%%%%%%%%%%%%

\section{The Uncertainty Principle}
How to make decisions under uncertainty? Decision-making under uncertainty relies mainly on how efficiently we can extract \textit{useful} knowledge from the data that were previously \textit{unknown} to the decision-maker\footnote{``Anything that gives us \textit{new} knowledge gives us an opportunity to be more rational'' – Herbert Simon}. C. R. Rao, in his \citeyear{rao1996} article\footnote{The existence of this article is barely known.} on `Uncertainty, Statistics, and Creation of New Knowledge' provided an exquisite
description of the mechanics of decision-making under uncertainty using a simple logical formula: 

\beq 
\setlength{\abovedisplayskip}{-1.25em}
\setlength{\belowdisplayskip}{1.5em}
\label{eq:rao}
\hspace{-.3em}\fbox{\text{Uncertainty of  knowledge}} ~+ ~ \fbox{\text{Knowledge of uncertainty}}\, =\, \fbox{\text{Usable knowledge.}}~
\eeq

A decision analyst confronts data $X_1,\ldots, X_n$ equipped with a tentative (imprecise and uncertain) probabilistic model $f_0(x)$ of the underlying phenomena. The challenge then boils down to effectively using the misspecified model $f_0(x)$ to learn from data and to apply that knowledge for informed decision-making. Rao's uncertainty principle suggests the following three-staged approach, which we call the `\textit{model-building triad}':
%For researchers, the challenge now is to  The challenge is to develop a theory of data analysis that can guide us 

%The key question becomes: How can we sensibly use an imperfect model $f_0(x)$ to learn from data for making informed decisions? 
%advocates
%Equipped with an imperfect probabilistic model, how to use data to guide decision-making? 
%Then the question is: How should an analyst use the misspecified model $f_0(x)$ to learn from the data for making informed decisions? 
% :
% Rao's uncertainty principle tells us that before making decisions based on the old model we have to perform two things: 

% principle

\vskip.75em

\texttt{Stage 1.} \textit{Model Elicitation}. The first step of decision making is formulating a probability model of the phenomena of interest, from which economic agents derive their initial expectations. A simple parametrized model-0 $f_0(x)$ is usually formed based on either gut instinct or the scientific context of the investigation. The \textit{uncertainty} of $f_0(x)$ arises due to the lack of perfect knowledge\footnote{what \cite{keynes1937general} called `uncertain knowledge.'} about the underlying probability law. Accordingly, the modeler has to start the analysis by acknowledging the uncertainty of the initial knowledge model $f_0(x)$.

% Due to the lack of perfect knowledge about the stochastic nature of a phenomenon

% (or sometimes to prioritize simplicity), we have inherent uncertainty or imperfection in the in $f_0(x)$. 

% we have inherent uncertainty or imperfection in the in $f_0(x)$ due to lack of perfect knowledge about the possible realizations of a phenomenon or sometime 

% Uncertainty arises from fuzzy knowledge, which is somewhere in between complete ignorance and perfect knowledge about the possible realizations of a phenomenon.

% The first step of How can a decision-maker sensibly use $f_0(x)$ by acknowledging misspecification.

\vskip.45em

\texttt{Stage 2.} \textit{Model Uncertainty Quantification}.  Before making decisions based on the provisional model $f_0(x)$, it is crucial to investigate its uncertainty (blind spots) in light of the new data. It's always a good practice to inspect expert opinions based on hard empirical facts by asking\footnote{Those who ignore experts' knowledge and only trust data are \textit{empirical-fools}. Those who ignore data and only trust their gut-instinct are \textit{emotional-fools} \citep{tversky1974}. Expert decision-makers always use empirically-guided intuition by \textit{appropriately} combining both data and available knowledge.}: what's new in the data that can't be explained by the assumed model? Discovering surprising and previously \textit{unknown} facts can prompt decision makers to consider other alternative actions.
%can improve decision makers' understanding about the matter.
%   Revealing
% \textit{surprising and previously unknown} facts helps  to the decision makers for a to   Have I overlooked something? Captures the \textit{surprising features} of the data.

\vskip.45em

\texttt{Stage 3.} \textit{Model Rectification and Risk Management}. Finally, we incorporate the learned uncertainty into the uncertain model $f_0(x)$ to produce a rectified model for  making empirically-guided informed decisions. It is important to \textit{sharpen} the ``judgment component'' (intuition based on past experiences) in light of the new data before it gets outdated.   

\vskip.75em

The purpose of this article is to describe a general statistical theory that permits us to implement this three-staged model-building procedure for data analysis and decision-making.

%to address the key question: How do we use imperfect model to learn from data?

% This article describes a theory of statistical learning theory with imperfect model that embraces and operationalizes Rao's uncertainty formula-based model-building principle. 
% that embraces and operationalizes Rao's uncertainty principle-based  model-building 

%Rao's uncertainty formula-based  three-staged model-building principle provides some systematic way of thinking about how to imperfect model to learn from data

%Rao's uncertainty formula

%law of model building 

%How to develop a statistical formulation that permits this principle to be applied for real applications?  This paper describes a simple and efficient method to actualize Rao's uncertainty formula model building principle. 

% Rao's equations provides a simple recipe: Uncertain model $f_0$ + knowledge of the amount of uncertainty in it = usable (useful + actionable) knowledge---based on which one can make decision.  

% is introduced that allows  

%%%%%%%%%%%%%%%%%%%%

\section{Learning with Imperfect Model}
\vskip1em
\begin{quote}
    \small{`\textit{All analysts approach data with preconceptions. The data never speak for themselves. Sometimes preconceptions are encoded in precise models. Sometimes they are just intuitions
that analysts seek to confirm and solidify. A central question is how to revise these
preconceptions in the light of new evidence}.'}
   \begin{flushright}
 \vspace{-.3em}
 {\rm --- \cite{heckman2017abducting}} \end{flushright}
\end{quote}
Empirical scientific inquiry typically starts with a simple yet believable model of reality (model-0) and aims to \textit{sharpen} existing knowledge by gathering new observations.
\vskip.25em
We observe a random sample $X_1,\ldots,X_n \mathrel{\dot\sim} F_0$. By ``$\mathrel{\dot\sim}$'' we mean 
that $F_0$ is an `approximately correct'  structured provisional model for $X$ that is given to us by subject-matter experts. We like to extract new knowledge from the data by smartly leveraging existing knowledge\footnote{Model amendment principle: the starting model $f_0(x)$ is incomplete but not useless. It contains valuable background knowledge. Rather than throwing this vital information, we want to build a model by smartly taking clues from it. The goal is to amend model-0, not to abandon it completely.} that is encoded in the initial approximate model $f_0(x)$.\footnote{As for notation: by $F_0(x)$, we denote the cumulative distribution function (cdf) of the starting model-0; $f_0(x)$ is the probability density function (pdf) and quantile function is denoted by $Q_0(u)$. The expectation with respect to $f_0(x)$ will be abbreviated as $\Ex_0$.} 
\vskip.25em
% $\bullet$ {\bf The Insufficiency Principle}. Based on this principle that.. Insufficient to completely describe the reality.  Inexactness of models; Models are always subject to revision in the light of new experiments and data; It is foolish to believe that the present model can not be replaced by some more exact model. 

% At same time they are ..misspecified but not useless..Rather than discarding them, can we use them is an intelligent way to build class of sensible models.  Models should be amended, not abandoned....$f_0$ provides a guidance in the form of `rough sketch' of the true model. 

% Thus, the goal should be to find a ``better'' model not the final model.

\textit{Creating knowledge-guided statistical models}. The core mechanism of our process involves: (i) inspecting whether the structured provisional model-0 is still a good fit in light of fresh data; (ii) if not, then we like to know what's new in the data that cannot be tackled by the current model; and, finally, (iii) repair the current misspecified model in order to cope with the new reality. However, the question remains as to how can we design an inference machine that can offer these successively fine-grained insights? 
To address this question, we will describe a new statistical model building principle, called the `density-sharpening principle.'

%Model misspecification: How do we use imperfect model to learn from data? Learning with Model Misspecification/imperfect model: 

% A scientific theory (say, economics, physics) the might suggest a model that is worthy to be tentatively entertained.

% {\bf Motivating question}. It requires at least three things: (i) Starts ...efficient way to check weather the current model is still a good fit for the new data environment. 

% To develop a theory of statistical learning, we have to design a computational program that mimics this process.  \textit{What is the purpose of a statistical model?} One of the primary goal is to discover \textit{new} knowledge for better understanding and decision making. 

% The goal is not to create a model for the observed data but to extract the `unexplained' part (containing the surprise) and build a model for that and finally, use it to rectify the initial model-0. How do we learn when presented with a new situation?

% which is data-consistent and at the same time ``as close as possible'' to $f_0$.

%  To this end, we propose some new nonparametric model refining principles and computational tools for transforming a crude initial model (Model-0) into a useful one (Model-1).

% by $X$ we denote a general (discrete, continuous, or mixed) variable with $F(x)$, mid-distribution function $\Fmn(x)=F_0(x) - \frac{1}{2}p_0(x)$. The associated quantile function will be denoted by $Q_0(u)=\inf\{x: F_0(x) \ge u\}$ for $0<u<1$.

%%%%%%%%%%%

\subsection{A Dyadic Model}

% typically coming from domain experts. typically guided by external background knowledge.  

We introduce a dyadic model with two interrelated subsystems that accommodates the decision maker's concern for misspecification of the starting expert-guided model.

\begin{defn}[Dyadic model] \label{def1:2tm}
$X$ be a general (discrete, continuous, or mixed) random variable with true unknown density $f(x)$ and cdf $F(x)$. Let $f_0(x)$ represents a simple approximate model for $X$ with cdf $F_0(x)$, whose support includes the support of $f(x)$.\footnote{For dealing with truly zero-probability events see \citet[Ch. 11]{coletti2002book}} Then the following dyadic density decomposition formula holds:
\beq \label{eq:fgd}
\setlength{\abovedisplayskip}{1.4em}
\setlength{\belowdisplayskip}{1.4em}
f(x)\,=\,f_0(x)\,d\big(F_0(x);F_0,F\big), \eeq
here $d(u;F_0,F)$ is defined as 
\beq 
\setlength{\abovedisplayskip}{1.4em}
\setlength{\belowdisplayskip}{1.4em}
d(u;F_0,F)= \dfrac{f(Q_0(u))}{f_0(Q_0(u))}, ~\,0<u<1,\eeq
\vskip.2em
where $Q_0(u)=\inf\{x: F_0(x) \ge u\}$ for $0<u<1$ is the quantile function. The function 
$d(u;F_0,F)$ is called `comparison density' because it \textit{compares} the initial model-0 $f_0(x)$ with the true $f(x)$ and it integrates to one:
\[\int _0^1 d(u;F_0,F)\dd u \,=\, \int_x d(F_0(x);F_0,F) \dd F_0(x) \,=\,\int_x \big(f(x)/f_0(x)\big) \dd F_0(x)\,=\, 1. ~~\]
\vskip.1em
However, we will interpret the $d$-function as the density-sharpening function (DSF), since it plays the role of ``sharpening'' the initial model-0 to hedge against its potential misspecification. To simplify the notation, $d(F_0(x);F_0,F)$ of eq. \eqref{eq:fgd} will be abbreviated as $d_0(x)$. 
\end{defn}

A few remarks on density-sharpening law: 
\vskip.35em
1. The model building mechanism of Definition 1  provides a statistical process of \textit{transforming and refining} a crude initial model into a useful one for better decision-making.
\vskip.45em
2. Note that if $d(u;F_0, F) \neq 1$, i.e., if $d(u;F_0, F)$ deviates from uniform distribution then \textit{change} of probability assignment is needed to embrace the current scenario. The density sharpening mechanism of \eqref{eq:fgd} prescribes how to revise the old probability assignments in light of new evidence. 
\vskip.45em
3. Similar to Rao's uncertainty law \eqref{eq:rao}, we can also write down a simple logical equation that captures the essence of the density-sharpening based model building principle (def. \ref{def1:2tm}):
\beq 
\setlength{\abovedisplayskip}{2em}
\setlength{\belowdisplayskip}{1.85em}
\label{eq:dslogical}
\hspace{-.45em}\fbox{\text{Misspecified model-0}} \,\,\times\, \,\fbox{\text{Knowledge of misspecification}}\, =\, \fbox{\text{Upgraded model-1}}
\eeq
\textit{Interpretation of the components}: the first component is the starting imprecise model $f_0(x)$, coming from expert knowledge. The second component $d_0(x)$ is the quality-assurer of the model that manages the risk of misspecification of the initial $f_0(x)$. $d_0(x)$ \textit{sharpens} the decision-makers initial mental model by extracting knowledge from data that
is previously unknown, which justifies its name---density sharpening function (DSF). Finally, the model-0 is ``stretched'' by $d_0(x)$ following eq. \eqref{eq:fgd} (only when the ideal scenario is different from the expected one) to incorporate the newly discovered information into the revised model. The class of $d$-sharp distributions turns the uncertain knowledge-distribution $f_0(x)$ into a usable distribution by properly sharpening using $d_0(x)$. Also, see Supp. A2, where 
a comparison between the traditional Bayes' rule and the density-sharpening-based multiplicative model update rule is presented.

\subsection{Comparison Coding}
The density-sharpening law provides a mechanism of building a model $f(x)$ for the data  $X_1,\ldots, X_n$ by \textit{inheriting} knowledge from the assumed working model $f_0(x)$. To apply the formula \eqref{eq:fgd}, we need to estimate $d_0(x)$ from data.\footnote{To keep the theory of estimation simple, we will mainly focus on the $X$ continuous case. A detailed account for the discrete case can be found in \cite{D21discrete}.} And we call this learning process `comparison coding' because $d_0(x)$ codes how surprising the current situation is in light of the model-0
by contrasting expectations with reality. 

% We \textit{learn by comparing} proposal density $f_0(x)$ (best guess model) and the unknown generative densities $f$, captured by the $d_0(x)$, which sharpens the shape of pre-specified $f_0$.

% $\bullet$ LP-basis for $X$ mixed. By $f_0(x)$ we mean the density function for the continuous case, and for the discrete case we mean the probability mass function $f_0(x)=\Pr_0(X=x)$.
\vskip.25em
Since the density-sharpening function $d_0(x):=d(F_0(x);F_0,F)$ is a function of $F_0(x)$, we can approximate it by a linear combination of polynomials that are function of $F_0(x)$ and orthonormal with respect to the base-model $f_0(x)$. One such orthonormal system is the LP-family of polynomials \citep{D20copula, D21discrete, Deep17LPMode}, which can be constructed as follows. For an arbitrary continuous $F_0$, define the first-order LP-basis function as \textit{standardized} $F_0(x)$:
\beq \label{eq:T1}
\setlength{\abovedisplayskip}{1.2em}
\setlength{\belowdisplayskip}{1.2em}
 T_1(x;F_0)\,=\,\sqrt{12} \big\{F_0(x) - 1/2\big\}. \eeq
Note that $\Ex_0(T_1(X;F_0))=0$ and $\Var_0(T_1(X;F_0))=1$. Next, apply Gram-Schmidt procedure on powers of the first-order LP-basis functions $\{T_1^2(x;F_0), T_1^3(x;F_0),\ldots\}$ to construct a higher-order LP orthogonal system $T_j(x;F_0)$:
\vspace{-1em}
\bea 
T_2(x;F_0) &=& \sqrt{5} \big\{ 6 F^2_0(x)  - 6 F_0(x) + 1 \big\}   \\
T_3(x;F_0) &=&  \sqrt{7} \big\{  20 F^3_0(x)  - 30F^2_0(x)  + 12F_0(x)     -1   \big\}   \\
T_4(x;F_0) &=&       \sqrt{9} \big\{ 70F^4_0(x)  - 140F^3_0(x)  + 90F^2_0(x)  -20F_0(x)     +1   \big\},
\eea
and so on. Compute these polynomials by performing the Gram-Schmidt process numerically, which can be done using readily available computer packages like R or python.

\vspace{.5em}
\begin{defn}[Comparison coding]
Expand comparison density in the LP-orthogonal series
\beq \label{cdm}
d_0(x) := d(F_0(x);F_0,F)\,=\,1+\sum\nolimits_j \LP[j;F_0,F] \,T_j(x;F_0).
\eeq
To estimate the unknown LP-Fourier coefficient, note that:
{
\addtolength\abovedisplayskip{-.1\baselineskip}%
  \addtolength\belowdisplayskip{-.3\baselineskip}%
\bea \label{eq:dlp}
\LP[j;F_0,F]&=& \int T_j(x;F_0) d_0(x) f_0(x) \dd x\\ \nonumber
&=& \int T_j(x;F_0) f(x) \dd x \\ \nonumber
&=& \Ex_F\big[  T_j(X;F_0)\big].
\eea
}
\end{defn}
Replacing $\LP[j;F_0,F]$ with its plug-in estimator in \eqref{cdm} we get
\beq \label{eq:dhat}
\setlength{\abovedisplayskip}{1.25em}
\setlength{\belowdisplayskip}{1em}
\wtd_0(x)\,=\,1+\sum\nolimits_j \tLP[j;F_0,F] \,T_j(x;F_0),
\eeq
where 
\beq \label{eq:eLPcoef}
\tLP[j;F_0,F] \,=\, \Ex_{\wtF}[T_j(X;F_0)] \,=\, \frac{1}{n} \sum_{i=1}^n T_j(x_i;F_0).\eeq

Although \eqref{eq:dhat} provides a robust nonparametric comparison-coding procedure, it has one drawback: the estimated $\wtd$ may be unsmooth due to the presence of a large number of small noisy LP-coefficients. To avoid unnecessary ripples in $\wtd$, we need to isolate the small number of non-zero LP-coefficients. Our denoising strategy goes as follows \citep{D21peirce}: sort the empirical $\tLP[j;F_0,F]$ in descending order based on their absolute value and compute the penalized \textit{ordered} sum of squares. This Ordered PENalization scheme will be referred as \texttt{OPEN} model-selection method:
\beq \label{eq:open}
\texttt{OPEN}(m)~=~\text{Sum of squares of top $m$ LP coefficients}~-~\dfrac{\gamma_n}{n}m.~~~~~~\eeq
Throughout, we use AIC penalty with $\gamma_n=2$. Find the $m$ that maximizes the ${\rm OPEN}(m)$. Store the selected indices $j$ in the set $\cJ$. The \texttt{OPEN}-smoothed LP-coefficients will be denoted by $\widehat{\LP}_j$. Finally, return the following \textit{smoothed} estimate: \beq \label{eq:sdhat}
\setlength{\abovedisplayskip}{1.15em}
\setlength{\belowdisplayskip}{1.15em}
\whd_0(x)\,=\,1+\sum\nolimits_{j \in \cJ} \hLP[j;F_0,F] \,T_j(x;F_0).
\eeq
\begin{rem}[The scientific value of sparse $d$]
The DSF $d_0(x)$ is the \textit{bridge} between the theoretical world (idealized model) and the empirical world (real observations).  A meaningful way to measure the simplicity of a model is the number of ``new'' statistical parameters that it contains \textit{beyond} the given scientific parameters---that is, the parsimony (number of parameters) of $d_0(x)$.\footnote{It selects only a handful of reasonable (rival) hypotheses out of a vast collection of possibilities. By `reasonable,' we mean hypotheses with a high \texttt{OPEN}($m$) score, which balances complexity (number of parameters of $d_0$) and accuracy (of explaining the surprising phenomenon).}  A sparse $\whd_0$ provides an intelligent and parsimonious way to elaborate the model-0 (not an indiscriminate, brute-force elaboration) to produce a `sophisticatedly simple' model. Simplicity is vital to make the model usable and interpretable by decision-makers, who like to understand \textit{how} to change the initial model to explain the data.
\end{rem}

% We are searching for ....  elaborate the class of distribution smarty (parsimoniously) not indiscriminately. 

%%%%%%%%%%%%%%%%%%%%%%%%%%%%%%%%%
\subsection{A Deep Dive into Model Uncertainty}
Understanding the deficiency of the current model is an essential part of the process of iterative model building and refinement: Have we overlooked something? Where are our knowledge gaps? This section provides a comprehensive understanding and exploratory tool for representing and assessing potential model misspecifications. Also, see Supp. note A1, which explains the distinctions between parametric and nonparametric model uncertainties.

%\footnote{See \cite{bankes1993exp} for an excellent discussion on the importance of exploratory modeling for decision-making and policy analysis.}

% measure of the incompleteness of the starting model. 

%What additional knowledge we can gain from the data that is new and previously unknown. 

%%%%%%%%%%%%%%%%%%%%%%%

\begin{figure}[ ]
\vspace{-.65em}
\includegraphics[width=.488\linewidth,keepaspectratio,trim=1cm 1cm 1cm 1cm]{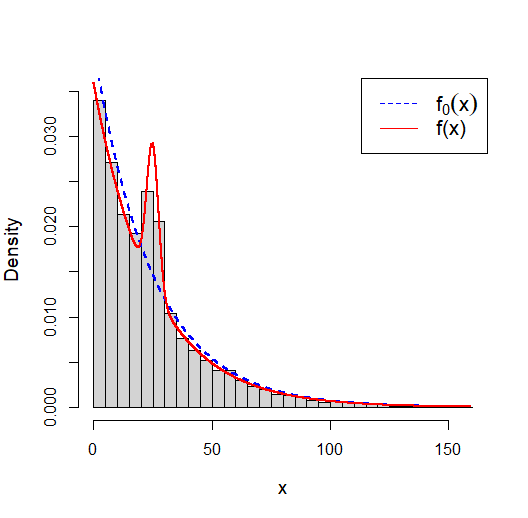}~~~
\includegraphics[width=.488\linewidth,keepaspectratio,trim=1cm 1cm 1cm 1cm]{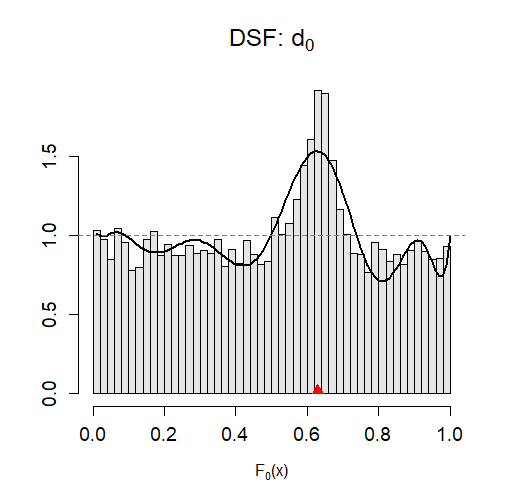}
\vspace{.65em}
\caption{How closely does reality resemble the model? $10,000$ samples are generated from the true (unknown to the analyst) model $0.9 {\rm Exp}(\la_0) + 0.1 \cN(25, 2.5^2)$. The right plot shows the graph of $\whd_0(x)$, which acts as a `magnifying glass' that forces us to examine what extra information data are willing to reveal \textit{beyond} the known assumed model. The crucial point here is: model uncertainty, as captured by $\whd_0(x)$, has a \textit{shape}---it is not a single numeric number. It helps physicists narrow down a large set of potential rival theories to a few plausible ones.} 
\label{fig:hep}
\end{figure}

\subsubsection{Graphical Exploration of Model Uncertainty} 

\begin{example}
\vspace{-.25em}
Consider the following scenario:  Fig. \ref{fig:hep} displays the data that a physicist just collected from an experiment. The blue curve is the physics-informed background distribution $f_0(x)$, which, in this case, is an exponential distribution with $\la_0=25$, and the red curve is the true unknown probability distribution. The physicist is mainly interested in knowing whether there is any \textit{new physics} hidden in the data, i.e., anything new in the data that was overlooked by existing theory. If so, what is it? How does the theory ($f_0$) relate to practice? This will help the physicist to come up with some scientific explanations and potential alternative theories.

\textit{The Shape of Uncertainty}. The researcher ran the density-sharpening algorithm of the previous section with $m=10$, and the resulting $\whd_0(x)$ is displayed in the right of Fig. \ref{fig:hep} as a function of $F_0(x)$. Few conclusions: (i) {\bf Model appraisal}: The non-uniformity of $\whd$ tells us that the ``shape of the data'' is inconsistent with the presumed model-0. (ii)  {\bf Model amendment}: The shape of $\whd$ also informs the scientist about the nature of deficiency of the old model---i.e., what are the most worrisome aspects of the presumed model? In this example, the most consequential unanticipated pattern is the presence of a prominent `bump' (excess mass) around $F_0^{-1}(0.63) \approx 24.85$, which might be indicative of new physics. This newly discovered pattern can now be used to improve the background exponential model. 
\end{example}

\begin{rem}[Visual explanatory decision-aiding tool]
One of the unique abilities of our exploratory learning is its ability to generate explanations on \textit{why and how} the model-0 is incomplete\footnote{Explanation-based statistical reasoning is at the core of abductive inference, as discussed later.}. Thus, the graph of $\whd(u;F_0,F)$ explicitly addresses decision-makers model misspecification concerns. It digs into the observations to uncover the ``blind spots'' of the current model that can ultimately drive discovery (locating novel hypotheses) and better decisions.   
\end{rem}

% \begin{rem}[MODEL SUPERVISOR]
% $d$ can be thought as  MODEL SUPERVISOR. which is responsible for checking whether $f_0$ is adequate, if not then it provides feedback/instructions how to rectify it (to resolve the discrepancy). It an important part of the model management toolbox. One can build increasingly SKILLED model as it gets to see more and more data. Density sharpening principle provides the revised probability model.
% \end{rem}

\subsubsection{Measure of Model Uncertainty}
A general measure of the \textit{degree} of model misspecification is defined using the Csisz{\'a}r information divergence class.

\vspace{.4em}

\begin{defn}
For $\psi: [0, \infty) \mapsto  \bbR$ a convex function with $\psi(1)=0$, define the Csisz{\'a}r class of statistical divergence measure between $F$ and $F_0$: 
\beq
I_{\psi}(F,F_0)\,=\, \int_{-\infty}^\infty \psi\left( \frac{f(x)}{f_0(x)} \right) f_0(x) \dd x\eeq 
We prefer to represent it in terms of density-sharpening function as follows: 
\bea 
I_{\psi}(F,F_0)&=& \int_{-\infty}^\infty \psi \circ d(F_0(x);F_0,F) \dd F_0(x) \nonumber ~~~~\\
&=& \int_0^1 \psi \circ d(u;F_0,F) \dd u,~~\text{where}~u=F_0(x).~~~~
\eea
\end{defn}
One can recover popular divergence measures by appropriately choosing the $\psi$-function:
\begin{itemize}[itemsep=8pt]
    \item KL-divergence: $\psi(x)=x\log(x)$; $I_{{\rm KL}}(F,F_0)=\int d \log d$.
\item Total variation divergence: $\psi(x)=|x-1|$; $I_{{\rm TV}}(F,F_0)=\int |d - 1|$.
\item Squared Hellinger distance: $\psi(x)=(1-\sqrt{x})^2$; $I_{{\rm H}}(F,F_0)=\int (1 - \sqrt{d})^2$.
\item $\chi^2$-divergence:  $\psi(x)=(x-1)^2$;  $I_{\chi^2}(F,F_0)=\int (d - 1)^2 = \int d^2 - 1$.
\end{itemize}
\vskip.4em

One can quickly estimate the $\chi^2$-model misspecification index by expressing it in terms of LP-Fourier coefficients (applying Parseval's identity to equation \ref{cdm}):
\beq \label{eq:chisq}
\setlength{\abovedisplayskip}{1.5em}
\setlength{\belowdisplayskip}{1.5em}
I_{\chi^2}(F,F_0)= \int d^2 - 1 = \sum_{j=1}^m \big| \LP[j;F_0,F] \big|^2.
\eeq 
$I_{\chi^2}(F,F_0)$ quantifies the uncertainty of the preliminary model $f_0(x)$ in light of the given data---i.e., whether $f_0(x)$ is catastrophically wrong or slightly wrong. Estimate it by plugging the empirical LP-coefficients \eqref{eq:eLPcoef} into \eqref{eq:chisq}. Since, under $H_0: F=F_0$, the sample LP-coefficients have the following limiting null
distribution (see Theorem 2 of \citealt{Deep17LPMode}):
\[ 
\setlength{\abovedisplayskip}{1.5em}
\setlength{\belowdisplayskip}{1.5em}
\sqrt{n} \tLP[j,F_0,F] \xrightarrow[]{d} \cN(0,1),~~ \textrm{i.i.d  for all}~ j,\]
$n \widetilde{I}_{\chi^2}(F,F_0)$ follows $\chi^2_m$ under null. One can use this to compute the $p$-value. Applying this measure to example 1, we get a $p$-value of practically zero---indicating that the background exponential model is badly damaged and should be \textit{repaired} before making a decision.

\begin{rem}
It is interesting to contrast our theory with \citet[Sec. 4 and 5.1]{hansen2022risk}, keeping in mind that their $m(x)$ is exactly  our sharpening function $d_0(x)$. This reinforces our belief that our theory can be applied broadly to econometrics and decision-making under uncertainty.
\end{rem}

%%%%%%%%%%%
\subsection{$d$-Sharp Models}
\vspace{.25em}
\begin{defn} \label{def:ds}
$\DS(F_0,m)$ stands for {\bf D}ensity-{\bf S}harpening of $f_0(x)$ using $m$-term LP-series approximated $d_0(x)$, given by: 
\beq  \label{DSm1}
\setlength{\abovedisplayskip}{1.3em}
\setlength{\belowdisplayskip}{1.5em}
~~~f(x)\,=\,f_0(x)\Big[ 1\,+\, \sum_{j=1}^m \LP[j;F_0,F]\, T_j(x;F_0)\Big],~~~~~
\eeq
obtained by replacing \eqref{cdm} into \eqref{eq:fgd}. 
$\DS(F_0,m)$ generates a relevant class of plausible models in the neighbourhood of the postulated $f_0(x)$ that are worthy of consideration. 

% This process generates a class of models in the vicinity of the postulated model 

%$\DS(F_0,m)$ is a self-improving class of models that adapt to new \textit{without being precisely told how beforehand.} 
\end{defn}

%Generates hypotheses and select the best one from the candidates. 

A few additional points on density-sharpening:
\vskip.14em

1. The $\DS(F_0,m)$-based density-sharpening principle provides a mechanism for \textit{exploring} data by \textit{exploiting} the uncertain background knowledge model. It starts with data and an approximate model $f_0(x)$---and produces a more refined picture of reality following \eqref{DSm1}.

%a mechanism that   

% It provides a seamless and consistent blending of theory-driven and data-driven models. Bridging the divide between theory and empirics.

% Here we define a `simple' model as the one that is close enough to the postulated scientific model, and at the same time, can explain the patterns in the data.

\vskip.24em

2. The process of density-sharpening suitably `stretches' the theory-informed model to create a class of robust empirico-scientific models. Moreover, it shows how new models are born out of data-driven \textit{mutation} of pre-existing ones.

% . =  A mechanism of enlarging the set of probability models
% by refining scientific knowledge based on empirical evidence.
\vskip.24em

3.  
The truncation point $m$ indicates the radius of the neighborhood around the elicited $f_0(x)$ to create permissible models. $\DS(F_0,m)$ models with higher $m$ entertain alternative models of higher complexity. However, to maintain conceptual appeal and interpretability, it is advisable to focus on the vicinity of $f_0$ by choosing an $m$ that is not too large. Substituting the smooth estimates $\hLP[j;F_0,F]$ of eq. \eqref{eq:sdhat} into the formula \eqref{DSm1}, we get the most economical model (among competing alternatives around $f_0$) that best explains the empirical surprise.\footnote{It brings our theory close to Gilbert Harman's ``Inference to the best explanation'' idea; see \cite{harman1965}. This is an area that merits further research.}

\vskip.24em
4. It provides an architecture of an `intelligent agent' that \textit{simultaneously} possesses the ability to: learn (what's new can we learn from the data), reason (how to explain the surprising empirical findings), and plan (how to self-modify to adapt in the new situations). 
% $\DS(F_0,m)$ class of models, which is often a central consideration when producing models that aid in efficient decision making.
% To avoid strange looking `spiky' alternatives: choose moderate $m$. provides a nice trade-off between conceptual appeal and
% computational tractability.

% construct families of models in the neighborhood of the assumed structured model

% large set of alternative models

% $\bullet$ The model building process starts by contrasting the exiting model-0 with data with the hope to learn something new, which is fully captured by $d_0(x)$. Then at the next stage it uses that newly discovered knowledge to build a more sharper model of reality, which in turn allow investigators to design more targeted future experiments.

% It provides a way to adapt the model: when the ideal scenario is different from the expected.

\vspace{.35em}

\begin{figure}[ ]
\vspace{-1.2em}
\includegraphics[width=.33\linewidth,keepaspectratio,trim=1.1cm 1cm 1cm 1cm]{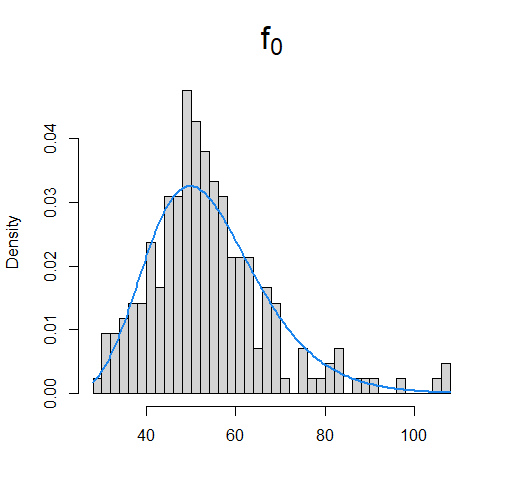}~~
\includegraphics[width=.33\linewidth,keepaspectratio,trim=1cm 1cm 1cm 1cm]{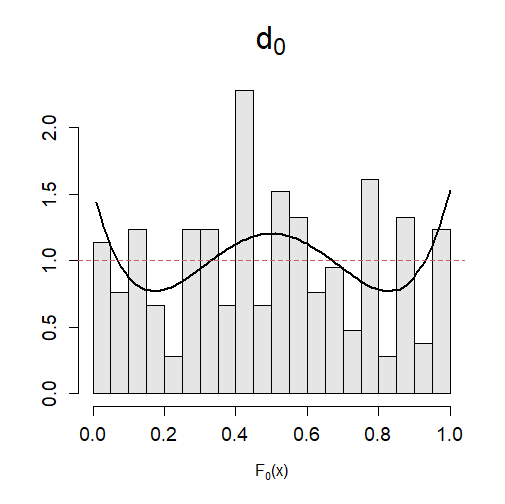}~~
\includegraphics[width=.33\linewidth,keepaspectratio,trim=1cm 1cm 1cm 1.1cm]{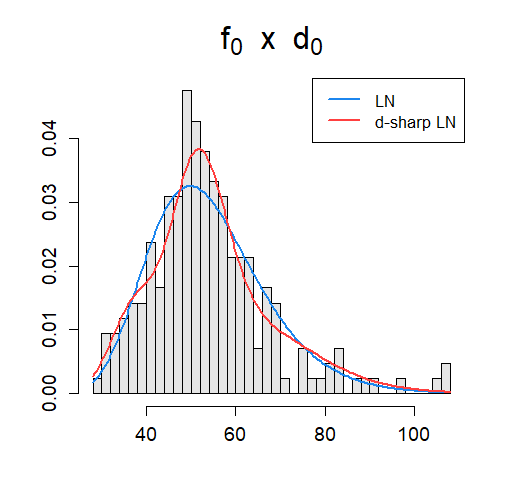}
\vskip.5em
\caption{Glomerular filtration data modeling. Left: the fitted lognormal distribution. Middle: The estimated density-sharpening function $d_0$ provides an economical description of the empirical surprise, thereby supplying clues for forming new explanatory hypotheses of the data. Right: the $d$-sharp lognormal with heavier tails and sharper peak.} 
\label{fig:grf}
\end{figure}

\begin{example}[Glomerular filtration data]
We are given glomerular filtration rates\footnote{Glomerular filtration rate (GFR) measures how much blood is filtered through the kidney to remove excess wastes and fluids. Low \texttt{gfr} value indicates that the kidneys are not functioning as well as they should.} for 211 kidney patients. The experiment was done at Dr. Bryan Myers' Nephrology research lab at Stanford University. The dataset was previously analyzed in \citet{efron2016computer}.

\vskip.33em
The blue curve on the left plot of Fig. \ref{fig:grf} shows the best-fitted lognormal (LN) distribution. We start our analysis by asking whether the parametric LN model needs to be refined to fit the data. The middle panel displays the density-sharpening function, which provides insights into the nature of misspecification of the LN model: the peak and the tails of the initial LN distribution need repairing; 
LN underestimates the peak and neglects the presence of heavier tails. The repaired LN model (displayed on right-hand side of Fig. \ref{fig:grf}) is given by
\beq \label{eq:gfr:fhat}
\setlength{\abovedisplayskip}{.7em}
\setlength{\belowdisplayskip}{.7em}
\hf(x)\,=\, f_0(x) \big[  1 \,+\, 0.18 T_4(x;F_0)  \big],
\eeq
where $f_0(x)$ is ${\rm LN}(\mu_0,\sigma_0)$, with $\mu_0=4$ and $\sigma_0=0.24$. The part in the square bracket comes from $d_0(x)$, which provides recommendations on how to suitably elaborate the LN-model to capture the \textit{unexplained} shape. The point of this example was to show how the density-sharpening principle (DSP) allows an analyst to explicitly perform model formulation, fitting, checking, and repairing---all seamlessly combined into one workflow.
\end{example}

It is interesting to compare our $d$-sharp LN-model (the red curve) with the seven-parameter exponential family fit shown in Fig.  5.7 of \citet{efron2016computer}. The most noticeable difference lies in the right tail. Efron's seven-parameter exponential family model shows weird spikes on the extreme-right tail. The main reason for this is that it is based on polynomials of raw $x$: ($x,x^2,\ldots,x^7$), which are not robust. That is to say, these traditional bases are unbounded and highly sensitive to `large' data points. In contrast, our LP-polynomials are functions of $F_0(x)$, not raw $x$, and thus robust by design. The other operational difference between our approach and Efron's exponential family approach is that we model the ``gap'' between lognormal and the data, which is often far easier to approximate nonparametrically (only required one parameter, see eq. \ref{eq:gfr:fhat}) than modeling the data from scratch.\footnote{There is an easy way to see that: compare the shapes of the histograms of the left two plots of Fig. \ref{fig:grf}.}

% -  It builds upon the theory-informed model $f_0$ and informs what other rival theories to be entertained. by properly sharpening the initial model in the light of the data.

% %%%%%%%%%%%%%%%%%%%%%%%%%%%%%%%%%
% \subsection{Exploratory Decision Analysis} 

% \textit{Exploratory Tool}.  Tools so that which aspects of the model-0 needs adjustment...Does it matter? ..can confront model
% misspecifications. imperfect model.. 

%%%%%%%%%%%%%%%%

% \begin{rem}[Model Architect]
% We need {\bf model architects}:  who can provide robust, efficient, and economical designs of models. facilitate the execution of building expansions or renovations.
% \end{rem}

%%%%%%%%%%%%%%%%%%%%%%%%%%%%%%%%%%%%%
\subsection{Modelplasticity and Abductive Inference Machine}
\begin{quote}
    \textit{Not the smallest advance can be made in knowledge beyond the stage of vacant staring, without making an abduction at every step. \hfill  {\rm --- C. S. \cite{peirce1901proper}}}
\end{quote}
{\bf Modelplasticity}---Models ability to modify and adapt itself in response to new data. The density-sharpening principle enables the model to develop \textit{new shapes} in the face of change.

%equipped with approximate knowledge of the underlying law. 

%model to adapt in novel situations and to and develop new shapes

% It provides the ability to adapt 

% Models are not fixed, its adaptable. It changes and evolves in response to new data.

% Explanatory: the root causes for doubting the competence of the model and also provides concrete prescriptions on how to modify the model leading to a new (hopefully an improved) tentative model---an \textit{iterative and interactive data discovery and model building.}

% Model adjust itself in order to be in concert with its surrounding environment (data). 

% Modelplasticity is an 

%

{\bf Density-sharpening and model evolution}. Modeling is a continual process, not a one-time data-fitting exercise.  The density sharpening mechanism allows us to combine new observations with a priory expected model to generate new insights, as depicted in Fig. \ref{fig:cycle}. 

\begin{center}
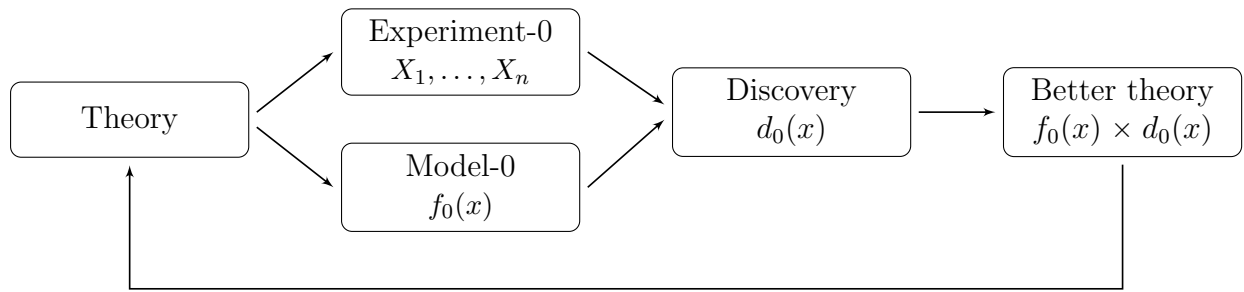
\begin{figure}[!h] 
\begin{tikzpicture}[node distance =4.4cm, auto]
    % Place nodes
    \node [block] (x1) {Theory};
    \node [block, right of    =x1,yshift=.9cm] (x21) {Experiment-0\\ {\small $X_1,\ldots,X_n$}};
    \node [block, right of    =x1,yshift=-.9cm] (x2) {Model-0 \\{\small $f_0(x)$}};
    \node [block, right of    =x21, yshift=-.8cm](x3){Discovery\\ $d_0(x)$};
    \node [block, right of    =x3](x4){Better theory\\ $f_0(x) \times d_0(x)$};
   
    \path [line] ($(x1.0)+(.1cm,-.1cm)$)--($(x2.180)+(-.1cm,0cm)$);
    \path [line] ($(x1.0)+(.1cm,.1cm)$)--($(x21.180)+(-.1cm,0cm)$);
    \path [line] ($(x2.0)+(.1cm,0cm)$)--($(x3.180)+(-.1cm,-0.1cm)$);
    \path [line] ($(x21.0)+(.1cm,0cm)$)--($(x3.180)+(-.1cm,0.1cm)$);
    \path [line] ($(x3.0)+(.1cm,0cm)$)--($(x4.180)+(-.1cm,0cm)$);
    \path[line] ($(x4.south)+(0,-.1)$) -- +(0, -4em) -| ($(x1.south)+(0,-.1)$);
\end{tikzpicture}
\vskip1.4em %Density-sharpening develops increasingly `better' model that explains new phenomena by broadening the scope of exiting knowledge base. 
 \caption{Architecture of abductive inference machine (\texttt{AIM}). Density-sharpening principle provides a systematic process of inserting the new information into the existing knowledge model to resolve empirical surprise and inconsistency. This continuous cycle of iterative model sharpening is called \textit{abductive learning}, which facilitates the emergence of new theories from data. As George \cite{box1980sampling} said, `The statistician's role is to assist this evolution.'}
\label{fig:cycle}
\end{figure}
\vspace{-3.2em}
\end{center}
{\bf Statistical law of model evolution}.  Density-sharpening supports this dynamic process of recursive model upgrading: $f_{k}(x) = f_{k-1}(x) \,d_{k-1}(x)$, for $k=1,2,\ldots$, by allowing the model to constantly evolve and reshape itself with fresh sets of data---going from a simple approximate model to a much more mature, accurate model of reality.

\vskip.2em
{\bf Abduction and creation of new knowledge}. Abduction is the creative part of an inferential process that aims at producing new theories from data. It builds upon what we know to discover new facts about nature. Abductive learning is concerned with the following questions: What \textit{new} can we learn from the data? 
How to \textit{change} the prior hypothetical model to explain the current situation? Which alternative classes of models are worthy of being entertained? 
\begin{quote}
    \small{`\textit{Does statistics help in the search for an alternative hypothesis? There is no codified statistical methodology for this purpose. Text books on statistics do not discuss either in general terms or through examples how to elicit clues from data to formulate an alternative hypothesis or theory when a given hypothesis is rejected.}'}
   \begin{flushright}
 \vspace{-.3em}
 {\rm --- C. R. \cite{rao2001statistics}} \end{flushright}
\end{quote}

Charles Sanders Peirce (1837–1914) was the pioneer of abductive reasoning; see \cite{stigler1978} and \cite{D21peirce} for more details on the Peircean view of statistical modeling. The goal of Abductive Inference Machine\footnote{They are not traditional pattern recognition (or matching) engine, they are  pattern \textit{discovery} engine.} or \texttt{AIM} is to provide a learning framework that endows a model with this ability to learn, grow and change with new information. 
\vskip.25em

% How to properly incorporate that to sharpen the exiting knowledge base? 
% Which alternatives should be considered before making a decision? 

% Abductive inference seeks to estimate the `most likely' probabilistic model (distribution) in the neighbourhood of $f_0(x)$ that could have generated the data.
% Where the uncertainty comes? (generation of alternatives) Because of the lack of one-to-one correspondence between data and hypothesis.
%  It provides a \textit{progressive} learning path `for growing knowledge and not pretending to have it' \citep{heckman2017abducting}.

\begin{rem}
The density-sharpening process plays an essential role for abductive inference, which provides the computational machinery for generating novel hypotheses with explanatory merit and selecting specific ones for further examinations.
\end{rem}

\begin{rem}[Abductive inference $\neq$ Hypothesis testing]
Any scientific inquiry begins with observations and some initial hypotheses. Classical statistical inference develops tools to test the validity of the null model in light of the data. Since all scientific theories are incomplete, accepting or rejecting a particular hypothesis is a pointless exercise. The real question is not whether the null hypothesis is true or false. The real question is: how far is the reality from the postulated model? In which direction(s) should we search to find a better model? Density-sharpening law provides a process of progressive refinement of yesterday's hypothesis.
\end{rem}

%%%%%%%%%%%%%%%%%%%%%%%%%%%%%%%

\subsection{Attention Mechanism}

\begin{quote}
    \textit{We often neglect how we get rid of the things that are less important...And oftentimes, I think that’s a more efficient way of dealing with information.} 
       \begin{flushright}
 \vspace{-.25em}
--- {\rm Duje Tadin\footnote{Jordana Cepelewicz (2019) To Pay Attention, the Brain Uses Filters, Not a Spotlight\textit{ Quanta Magazine}, https://www.quantamagazine.org/to-pay-attention-the-brain-uses-filters-not-a-spotlight-20190924.}}~~\end{flushright}
\end{quote}
Attention is the prerequisite of gaining new knowledge. Intelligent learners have the ability to quickly \textit{infer} where to focus attention to gain knowledge. In our modeling framework $d_0(x)$ draws analyst's attention quickly and efficiently to the new informative part by suppressing boring details; verify it from the graphs of $d_0(x)$ in Figs. \ref{fig:hep} and \ref{fig:grf}. It acts as a `gating mechanism' that filters out the new interesting (surprising) aspects of the data, and ignores the dull and unsurprising part---thereby sharpening the model's intelligence by guiding where to pay attention for information processing.

% The idea of `Learning by comparison' encapsulates the attention mechanism within it. 

% $d_0(x)$ makes it an attentive model..

%Establishing priorities 

% this helps the model to keep sustained attention on the `surprising' part of the data that need focus.

% distracts model's attention from the irrelevant information. higher-level  

% \begin{rem}
% It has some interesting connection with Herbert Simon's `difference operator' that learns and represents the `\textit{difference}' between the desired and present model from the data; see \cite{D21peirce} for more details.  
% % searchlight: filters out the least interesting part of the data. 
% \end{rem}

% suppresses boring information and amplifies only the new aspects (knowledge). Its like a filter. (We attend to only a fraction of the sensory data available to us; brain filters out the sensations least interesting to it at any moment.) Neuroscientists want to determine the circuits that aim and power that searchlight.

% provides the necessary attentional abilities---enabling the model to selectively concentrate on the interesting (surprising) part of the data, ignoring the dull part. 

% Our density-sharpening based two-system model: disentangle known from unknown to pay attention. system-2 $d$ stores the `new part.'
% $d$ performs comparison and the NEW information is then passed to the next-level higher-order information processing ($d$) where computational happens. $d$ makes the model \textit{attentive} to new information by ignoring boring.

% HEP example: highlight how $d$ removes the boring part and focuses into the new.

\vspace{-.5em}
 \begin{quote}
    ``\textit{The whole function of the brain is summed up in: error-correction}'' \vskip.24em
    \hspace{.8in}--- {\rm W. Ross Ashby},  English psychiatrist and a pioneer in cybernetics.
\end{quote}

\begin{rem}
In the brain, a dedicated circuit (or system) performs information-filtering similar to what $d_0(x)$ does for our dyadic model. The existence of such a brain circuit was first hypothesized by Francis \cite{crick1984function}---he called it `The Searchlight Hypothesis.' Since then, significant progress has been made to hunt down the brain region, what is now called basal ganglia,  that suppresses irrelevant inputs. For more details see \cite{halassa2017} and \cite{gu2021com}. Basal ganglia help us focus on what's important and tune out the rest. The mechanics of our model-building mimic the brain's cognitive process that uses existing knowledge to sieve out the new information for correcting the error (sharpening) of the earlier mental model. 
\end{rem}

\section{Decision-Making with Imperfect Model}

%%%%%%%%%%%%%%%%%%%%%%%%%
\begin{quote}
\textit{How should a decision maker acknowledge model misspecification in a way that guides the use of purposefully simplified models sensibly?}
   \begin{flushright}
 \vspace{-.25em}
 {\rm --- \cite{larscerreia2020}} \end{flushright}
 \end{quote}
 \vspace{-.1em}
This section demonstrates how practicing abductive inference based on the density-sharpening principle can enable better decision-making in highly uncertain environments.

\subsection{Abductive Model of Decision Making}
Abduction is the process of generating and revising a model \textit{before} choosing the optimal action. An abducer makes decisions in a dynamic uncertain environment by allowing for potential model misspecification.\footnote{The importance of model uncertainty in economics, finance, and business is beautifully illustrated in \cite{hansen2014}, although from a different perspective.} Abductive decision-making is about knowing \textit{when} to change course and \textit{how} to change it.  

%provides a precise and richer description of reality
\vskip.35em
How can a decision-maker \textit{abduct}? The mechanics of abductive decision-making consist of three steps: (i)  generating a set of plausible alternative models based on new evidence; (ii) constructing a `robust' model (by choosing the least favorable alternative model or by averaging the alternative models with proper weights); and (iii) selecting an action that maximizes expected utility under the newly revised model. Two modes of abductive decision-making under uncertainty are presented below.

\vskip.35em

{\bf Notation}. A decision-maker (DM) has to take an action $a$ from the set of available actions $\mathbb{A}=\{a_1,\ldots,a_q\}$ based on observed outcome $X_1,\ldots,X_n$ from an unknown probability distribution $f(x)$, representing  some natural or social
phenomenon. The DM selects the optimal action that minimizes expected loss (or risk) under the assumed model-0:
\beq \label{eq:dec1}
\ha_0 := \argm_{a \in \Abb} \int L_a(x) \dd F_0(x),
\eeq
where $f_0(x)$ is the DM's posited probability distribution over outcomes.  However, as an abducer, the DM is completely aware that the uncertainty about the outcomes may not be fully captured by a single, rigidly-defined probability distribution $f_0(x)$ and thus wants to choose the best decision by accommodating the uncertainty of model-0.

% $\bullet$ Uncertainty refers to the ``gap'' between the available knowledge (which is 
% plagued by imperfect and incomplete knowledge about the stochastic phenomena) and the requisite knowledge to make the best policy choice----which is captured by $d(u;F_0,F)$.

\vskip.35em
{\bf Decision making based on density sharpening principle}.  
To account for the imperfect nature of model-0, the most natural thing to do is to work with an enlarged class of plausible distributions around the vaguely acceptable $f_0(x)$:
\beq  
\Gamma_M~=~\big\{  f: \, f \in \DS(F_0,m),~ m \le M  \big\}
\eeq
within a certain reasonable neighbourhood, say $M=10$. We like to use this enlarged class of distributions $\Gamma_M$ for robust decision-making. Two such strategies are discussed below.
%How to use this enlarged class of distributions $\Gamma_M$ for robust decision making? 
% Decision makers, being afraid that the reference model $f_0$ might not be correct, also entertain to decide best possible actions .

% To make better decisions in a changing world, the decision system has to be driven by a model that can continuously learn and grow.

\vskip.35em
{\bf Method 1.} A cautious DM selects an action by its expected loss under the least favourable distribution within the set  $\Gamma_M$:
\beq \label{eq:dec2}
\breve{f}_{a,M}\, =\, \argsup_{F \in \Gamma_M} \int L_a(x) \dd F(x).
\eeq
We call this an \textit{abductive-minimax} procedure. Our proposal is partly inspired by the `local-minimax' idea of \cite{hansen2001, hansen2001robust}.

% This is in sharp contrast with Classical decision theory \citep{wald1950}, where it is DMs pretend to know the true probability model. 

% We introduce a formal mathematical language for implementing this 

% Our learning architecture allows an `analyst to engage models and data in an iterative dynamic process' \citep{heckman2017abducting}.

\vskip.35em
{\bf Method 2.} We now describe another robust decision-making procedure that takes into account the uncertainty in the analyst's elicited probability model of future states. Two key concepts are: bootstrap model averaging and action-profile function.
\vskip.25em
~\texttt{Step 1.} We use bootstrap to explore $f \in \Gamma_M$ in an intelligent way. Draw $n$ samples with replacement from the original data. Denote the bootstrap empirical cdf as $\wtF_*^{(1)}$. Perform density-sharpening algorithm based on $\wtF_*^{(1)}$, and denote the estimated $d$-sharp model as $f_*^{(1)}$.

~\texttt{Step 2.} Use $f_*^{(1)}$ to select the best action from the given set of $q$-actions $\{a_1,\ldots, a_q\}$. Denote the selected action as $a_*^{(1)}$.

\begin{figure}[ ]
\vspace{-1em}
\centering
\includegraphics[width=.77\linewidth,keepaspectratio,trim=2cm 2cm 2cm 2cm]{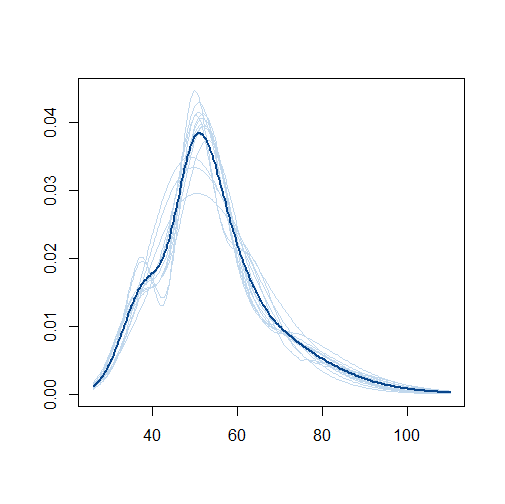}
\vskip.5em
\caption{The light blue curves are the bootstrap generated gfr densities, which try to present a landscape of plausible scenarios as suggested by the data. The dark blue denote the estimated model-averaged distribution $\bar f(x)$, which takes into account the uncertainty surrounding the  initially postulated model $f_0(x)$ as captured by the class $\Gamma_M$.} 
\label{fig:bma}
\end{figure}

~\texttt{Step 3.} Repeat steps 1-2, $B$ times (say $B=1000$ times). And return: 
\begin{itemize}[itemsep=4pt]
    \item The sample bootstrap \textit{distribution} $p_A$ of optimal actions $\{a_*^{(1)}, \ldots, a_*^{(B)}\}$---which we call the \textit{action profile} of the decision problem. 
    \item Bootstrap systematically generates probable alternative models $\{f_*^{(1)}(x),\ldots, f_*^{(B)}(x)\}$ from the class $\Gamma_M$ that can explain the data. Compute bootstrap model averaged distribution\footnote{This is also known as bagging \citep{breiman96bagging} or bootstrap smoothing \citep{efron2014est}.}:
\beq 
\label{eq:fbar}
\bar f(x)\,=\, \frac{1}{B}\sum_{j=1}^B f_*^{(j)}(x).~~~~~
\eeq
By averaging all plausible alternatives, $\bar f(x)$ becomes robust to model uncertainty. In this strategy, the policymaker does not have to put his/her complete faith in a single alternative distribution to assign probabilities. Bootstrap density exploration generates and weights different alternatives (from the class $\Gamma_m$) appropriately to create a realistic model. Fig. \ref{fig:bma} shows the bootstrap-generated densities for the \texttt{gfr} data of example 2. The light blue curves are the plausible alternative models, and the dark blue is the averaged density that takes into account all likely scenarios. 
It's worth contrasting $\bar f(x)$ with more traditional parametric uncertainty-based Bayesian predictive density; see supplementary A1.
\end{itemize}

\begin{rem} 
Two remarks: (i) Rather than assuming that alternative probable models are given to us \textit{a priori}, we use density-sharpening-based abductive inference method to \textit{synthesize} them, allowing us to account for a much broader range of model uncertainty than is possible with conventional approaches like Bayesian model averaging; see Supp. note A4. (ii) Furthermore, in our scheme, bootstrap provides a way to \textit{automatically estimate} the posterior probability of accepting each synthesized alternative model (light blue curves in Fig. \ref{fig:bma}), eliminating the need for arbitrary subjective probabilities over models. To compute the $\bar f$, bootstrap-derived model posterior probabilities are used to weigh the various alternatives from the density-set $\Gamma_M$; see Supp. note A3.
\end{rem}

~\texttt{Step 4.} This bootstrap scheme can also be used to approximately compute the least favourable distribution, defined in \eqref{eq:dec2}:
\[ \breve{\breve{f}}_{a,M} = \argmax_j \int L_a(x) \dd F_{*}^{(j)}(x).\]
The decision-maker can use this estimated model to carry out the proposed abductive-minimax procedure (method 1).

% Acidity index data: Select and display $5$ bootstrapped distributions and also display the $\bar f(x)$. Based on the cleanliness of the water the policymakers make take different actions like changing water source, or repairing the the whole infrastructure of aging pipes etc.

~\texttt{Step 5.}  {\bf Robust procedure}\footnote{Our philosophy of robustness is in complete agreement with \cite{huber77robust}, who advocated \textit{distributional} robustness: ``one would like to make sure that methods work well not only at the [idealized parametric] model itself, but also in a neighborhood of it.’’}: A pragmatic\footnote{Pragmatism is the logic of abduction.} decision-maker chooses an action (or ranks the actions) that minimizes expected loss (or maximizes the expected utility) with respect to the averaged-distribution: $\ha_{{\rm robust}} := \argm_{a \in \Abb} \int L_a(x) \dd \bar{F}(x)$. Our strategy prescribes action that is robust across a wide range of plausible alternative models. It could be especially powerful for dealing with ``deep uncertainty'' in making robust policies.  For a comprehensive overview on this subject, see \cite{marchau2019decision}. 

% \beq \label{eq:decR}
% \ha_{\DS} := \argm_{a \in \Abb} \int L_a(x) \dd \bar{F}(x)
% \eeq

~\texttt{Step 6.} Quantifying the `robustness' of the action (or decision rule):  How much does the optimal action change when a model is selected from a reasonable neighborhood  of the assumed initial opinion,  i.e. from the $\Gamma_m$ class? The shape of the action profile distribution can be used to determine how robust the optimal action is to model perturbation. In particular, the entropy of the action profile distribution can be used to  assess the robustness (or stability) of the inference to potential model misspecification: 

\beq 
{\rm Entropy}[p_A] \,=\,- \sum_{i=1}^q p_A(i) \,\log p_A(i) = \,- \sum_{i=1}^q \Pr(A=a_i)  \,\log \Pr(A=a_i).
\eeq
Uniform probability over possible actions yields maximum uncertainty---indicating that the decision is highly non-robust (sensitive) to model misspecification.

% Another option: decisionmaker chooses the alternative that maximizes the minimum expected utility across the possible alternative models  ...is clearly an alternative to the standard expected utility model. whatever probability
% model a policymaker might have, it cannot be known with certainty. Considering a
% set of priors around a given model and asking how robust economic policy would
% be to variations in the underlying probability

% APF:  allow the decision maker to
% assess formally the potential consequences of misspecification

%%%%%%%%%%%%%%%%%%%%%%%%%%%%
\subsection{Quantile Decision Analysis}
% Topic is \textit{Expert Model Elicitation}.
Until now, we have assumed experts can precisely formulate their opinion in a probabilistic form $f_0(x)$. However, for complex real-world decision-making problems, experts might only have \textit{incomplete} information about the uncertainty distribution of the target variable. A decision-analyst often elicit their partial knowledge about an uncertain quantity as a set of quantile-probability (QP) pairs $\{x_i, F(x_i)\}$, for $i=1,\ldots,\ell$. The job of an analyst is to find a simple, flexible, and parameterizable density that honors the assessed percentiles.  
\vskip.3em
% Based on a small set of QP pairs, analysts want a complete description of the expert’s distribution.

% experts probability distribution  

% Incomplete information: The most common scenario: An analyst obtains information about the expert’s distribution for X by
% assessing several QP pairs 

%  ,..In this case, a decision analyst (or analyst) has assessed an expert's $\{0.1, 0.5, 0.9\}$ quantiles..expert’s knowledge from a set of given QP pairs.
\vskip.25em

{\bf Model ambiguity due to incomplete information}. The task of eliciting an expert's probability distribution from a small set of QP pairs is a vital yet nascent topic in decision analysis; see \cite{powley2013quantile, keelin2011quantile, hadlock2017quantile}. In this section, we present an algorithm called \texttt{Q2D} (stands for quantile to distribution) that provides a systematic approach to deduce a reliable expert distribution from $\ell$ arbitrary QP-specifications. 
% The proposed \texttt{Q2D} algorithm provide a  using quantile judgments to deduce a probability distribution that reflects expert’s partial knowledge. 

% \vskip.5em
% :
% it's simple (mathematically tractable with explicit form) to interpret and honor  the 
% The task is to deduce a probability distribution that honor (at least approximately) expert’s quantile judgments. This 
% %Based on a small set of QP pairs

\vskip.35em
{\bf Probability-gap Approximation}. The main theoretical idea behind \texttt{Q2D} algorithm: Recall our $\DS(F_0,m)$ model
\beq f(x)\,=\,f_0(x)\Big[ 1\,+\, \sum_{j=1}^m \LP[j;F_0,F]\, T_j(x;F_0)\Big]~~~  \eeq
Integrating from minus infinity to $x$ on both sides, we have
\[ \int_{-\infty}^x ( f(z) - f_0(z) ) \dd z =  \sum_{j=1}^m \LP[j;F_0,F] \int_{-\infty}^x S_j(F_0(z)) \dd F_0(z),\]
where $S_j(u)=T_j(Q_0(u); F_0)$ is defined over the unit interval $[0,1]$ and $Q_0(u)$ is the quantile function of the distribution $f_0$. This leads to
\beq \label{eq:qe1}
F(x) - F_0(x)\,=\, \sum_{j=1}^m \LP[j;F_0,F] \int_0^{F_0(x)} S_j(u) \dd u.
\eeq

Given a set of arbitrary $\ell$ quantile-probability data $(x_i,F(x_i)),$ for $i=1,\ldots,\ell$, we can rewrite \eqref{eq:qe1} compactly as a matrix equation
\beq 
v = S_0 \bbe
\eeq
where $v_i=F(x_i) - F_0(x_i)$, $\be_i=\LP_j$, and $S_0 \in \cR^{\ell\times m}$, $S_0[i,j]=\int_0^{F_0(x_i)} S_j(u)$. The desired parameters are ${\bm \be}=(\be_1,\ldots,\be_m)$, where $\be_j$ is shorthand for $\LP[j;F_0,F]$.

For $m \le \ell$, we can uniquely estimate $\bbe$ using the least-square method
\beq 
\widetilde{\bbe}\,=\,\miniz_{\bbe} \| v - S_0 \bbe \|^2 \, =\, (S_0^TS_0)^{-1}S_0^{T}v.
\eeq
For large $\ell$ (say, $\ell\ge 5$), a better, more stable estimate can be found through regularization
\beq \label{eq:lasso}
\widehat{\bbe}\,=\,\miniz_{\bbe} \| v - S_0 \bbe \|^2~+~\la \|\bbe\|_1
\eeq
where $\| \cdot\|_p$ is the $\ell_p$ norm, and $\la>0$ is the regularization parameter.
The lasso \citep{lasso1996} penalized  $\widehat{\bbe}$ yields a sparse estimate and counters over-fitting. This penalized estimate provides a tradeoff between accuracy and interpretability.\footnote{Note that due to regularization, Eq. \eqref{eq:lasso} can even tackle cases with $m > \ell$. In such scenarios, the OLS is ill-posed, with an infinite number of solutions.} Finally, plug the estimated LP-Fourier coefficients $\be_j$ 
into the primary equation \eqref{DSm1} to get the expert distribution. 

%Note that expert quantile specifications might carry `judgmental error' or `hindsight bias.'
\begin{rem}
The expert quantile specifications should not be viewed as a `gold standard'---they are nothing but a preliminary guess (prone to errors of judgment or hindsight bias) whose purpose is to steer the analyst in the right direction\footnote{\cite{winkler1967} emphasized that the expert does not have some `true' density function waiting to be elicited, only a `satisficing' initial distribution that the policymaker is `content to live with at a particular moment of time.'}. For that reason, we recommend the smoothed (denoised) regularized $\widehat{\bbe}$ over the naive $\widetilde{\bbe}$, since it makes little sense to find an exact fit to the noisy QP-data.  
\end{rem}

% \begin{rem}
%  If we accept
% this position, it follows that there can be no unique ‘satisficing’ prior distribution..we doubt whether an analyst could ever claim that a single density function $g$ was the sole
% correct representation of an expert’s beliefs'
% \end{rem}

\begin{figure}[ ]
\vspace{-.7em}
\includegraphics[width=.46\linewidth,keepaspectratio,trim=.7cm .7cm 0cm 0cm]{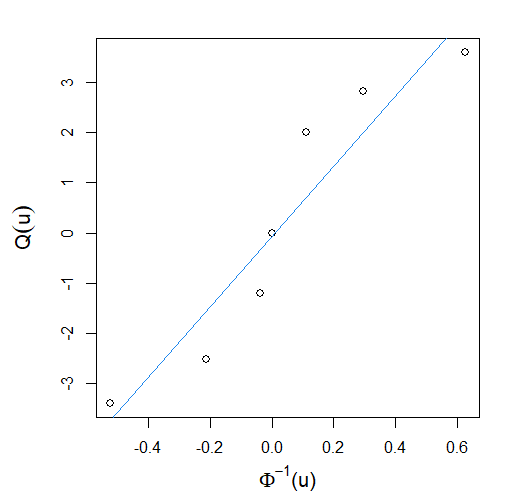}~~~~
\includegraphics[width=.465\linewidth,keepaspectratio,trim=1cm 1cm 1cm 2.5cm]{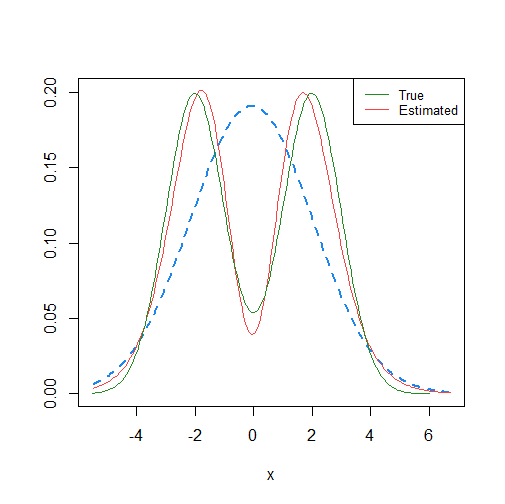}
\vskip.55em
\caption{Learning from incomplete information. Left panel: Regression on the scatter $(\Phi^{-1}(u_i), Q(u_i))$, $i=1,\ldots,7$. Right panel: The blue dotted line is the estimated $f_0$: normal distribution. \texttt{Q2D} estimated curve is shown in red, which is impressively close to the true mixture density $\frac{1}{2} \cN(-2,1) + \frac{1}{2} \cN(2,1)$.} 
\label{fig:bimodal}
\end{figure}
\vspace{-.6em}
\begin{example}[Bimodal Distribution]
We are given the following quantile judgments:

\begin{table}[h]
\vskip.3em
\centering
\begin{tabular}{l | ccccccc}
 Quantile: $x_i$ & -3.40 & -2.53 & -1.20 & 0 &  2.0&  2.83& 3.60 \\[.4em]
 \hline & \\[\dimexpr-\normalbaselineskip+.7em]
Probability: $F(x_i)$ & 0.04 &0.15& 0.39& 0.50 &0.75 &0.90& 0.97
\end{tabular}
\end{table}

In our \texttt{Q2D} algorithm, we choose $F_0$ (an initial approximate shape) to be normal distribution. To estimate the  parameters $\mu_0$ and $\sigma_0$ of the normal distribution, note that the quantile function $Q(u) \approx \mu_0 + \sigma_0 \Phi^{-1}(u)$. Thus one can quickly get a rough estimate by simply performing a linear regression\footnote{This technique will work for any location-scale family $f_0(x)$, e.g. normal, Laplace, logistic, etc.} on $(\Phi^{-1}(u_i), Q(u_i))$; see Fig. \ref{fig:bimodal}. The estimated normal distribution is shown in the right panel, along with the \texttt{Q2D}-estimated density.
\end{example}

\begin{figure}[ ]
\includegraphics[width=.33\linewidth,keepaspectratio,trim=1.1cm 1cm 1cm 1cm]{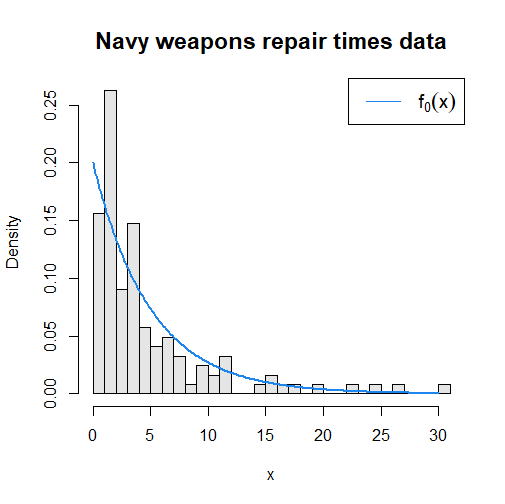}~
\includegraphics[width=.32\linewidth,keepaspectratio,trim=1cm 1cm 1cm 1cm]{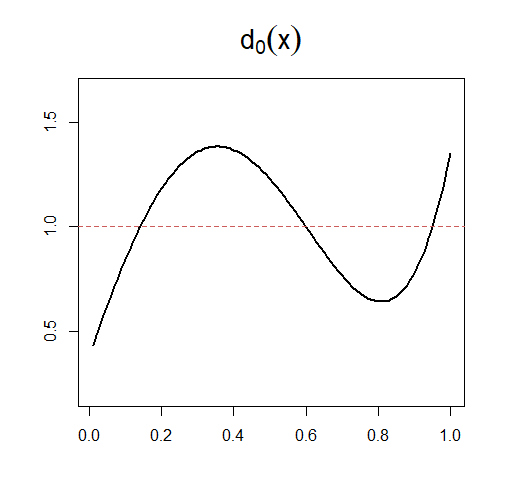}~~~~
\includegraphics[width=.34\linewidth,keepaspectratio,trim=1cm 1cm 1cm 1.1cm]{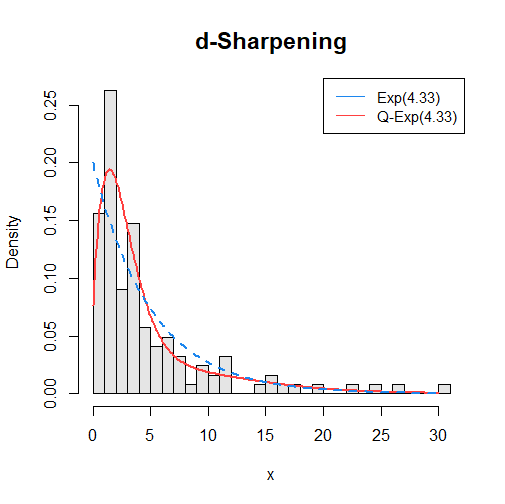}
\vskip.85em
\caption{Left: histogram of 122 repair times for a component of a U.S. Navy weapons system. The blue curve is the $f_0={\rm Exp}(4.32)$. Our analysis only used five QP-data (not the full data), whose outputs are shown in the middle and right-hand panels. The inferred density-sharpening function tells that the peak and the tail of the exponential model need correction. The repaired exponential model is displayed in red.} 
\label{fig:navy}
\end{figure}

%%%%%%%%%%%%%%%%%%%%%%%%%%%%%%%%%%%%
\begin{example}[U.S. Navy data]
Fig. \ref{fig:navy} shows a histogram of 122 repair times (in hours) for a component of a U.S. Navy weapons system. The dataset was analyzed in \cite{law2011select}. Imagine that for privacy and other reasons, we do not have access to the full data. The goal is to infer a probability distribution that faithfully represents the following quantiles:
\begin{table}[h]
\vskip.3em
\centering
\begin{tabular}{l | ccccc}
 Quantile: $x_i$ & 0.12 & 1.30&  3.00&  7.00& 26.17  \\[.4em]
 \hline & \\[\dimexpr-\normalbaselineskip+.7em]
Probability: $F(x_i)$ & 0.01 &  0.20&  0.50&  0.80&  0.99 
\end{tabular}
\end{table}

We start with exponential distribution as our initial guess, which is often taken as a `default' distribution (model-0) in reliability analysis. For $X \sim {\rm Exp}(\la)$, we have 
\[{\rm Median}(X)\, =\, \la \ln(2),~~{\rm where}~ \la=\Ex(X).~~\]
From the quantile table we get $\hat \la = 3/\ln(2) = 4.32.$ Next, we apply the \texttt{Q2D} algorithm to derive the LP-parameters with $f_0={\rm Exp}(4.32)$. The resulting density sharpening function and the final $d$-sharp exponential are shown in Fig. \ref{fig:navy}. The red curve on the right plot shows an excellent fit to the data, which was derived by the \texttt{Q2D} algorithm simply by utilizing the five quantile-probability pairs. 
\end{example}

% {\bf Scenario 2.} In addition to the quantile specifications, we are also given a small number of handful of data. Use \texttt{Q2D} algorithm to derive a reasonable starting $f_0(x)$ and at the second stage, use DS method to check its validity and refine it further based in the light of the given data.

% Rather than producing a
% single $f_0$, the analyst needs to derive a set of all plausible density functions that he
% believes to be consistent with the judgements provided by the expert and the data at hand.

%%%%%%%%%%%%%%%%%%%%%%%%%%%%

% An approximate expert’s distribution.

% QP pairs are obtained by asking the expert questions like: “What is your2 probability that X is less than 20?”; or
% “What value of X is so small, that you believe there is only a 10\% chance that the realized
% value for X will be less than this number?”

\vspace{-.88em}

%%%%%%%%%%%%%%%%%%%%%%%%%%%%
\subsection{Decision-making based on Multiple Experts}
High-stakes decision-making  (say, COVID-19 pandemic or climate change) is often based on multiple experts' opinions instead of putting all bets on a single rigidly-defined probability model. The challenge is to aid data-driven decision-making by appropriately combining several experts' models. We describe one possible way to build a `\textit{consensus committee model}' that can be used as a possible model-0 within an abductive decision-making framework.

\vskip.4em

{\bf Learning from multiple expert distributions}. Given $k$ competing probability models $\{f_{01},\ldots, f_{0k}\}$, which may differ markedly in shape, define the following model-weights:
\beq 
\text{Relevance weight:}~~~~~~w_\ell = \dfrac{1}{1 + \sum_j |\LP_{j|\ell}|^2},~~~~{\rm for}~ \ell=1,\ldots,k ~~~~~~~~~~\eeq
where $\LP_{j|\ell}$ is the LP-Fourier coefficients of the $\ell$-th model:
\beq 
d_{0\ell}(x) := d(F_{0\ell}(x); F_{0\ell},\wtF)\,=\, 1+\sum_j \LP_{j|\ell} T_j(x;F_{0\ell}).
\eeq
Note that the \textit{relevance weight} for the $\ell$-th model is always $0 < w_\ell \le 1$, and 
\[ w_\ell=1 ~~~\text{if and only if}~~~\LP_{j|\ell}=0,~\forall~ j.~~ \]
$\LP_{j|\ell}=0$ for all $j$ when $f_{0\ell}$ fully explain the data and there is no need to sharpen it further (i.e., $d_{0\ell}=1$). In that sense, $w_\ell$'s are data-driven weights (which will keep changing as we get more and more fresh data), computed based on the degree of agreement between the observed data and expert model $f_{0\ell}$. Define mixture expert distribution as 

\beq 
f^0_{\rm mix} (x)~= ~\sum_{\ell=1}^k \pi_\ell\,  f_{0\ell} (x),~~~\eeq
where $\pi_\ell = w_\ell/\sum_\ell w_\ell$. This model serves two purposes: it tries to resolve conflicting opinions based on data and at the same time encourages one to include as much diverse information as possible. 

\textit{Additional layer of model uncertainty}. If the analyst believes that the correct model might not be among the collection of models being considered, then use the combined expert model $f^0_{\rm mix} (x)$ as a model-0 in the subsequent density-sharpening-based learning and abductive decision-making process.

%%%%%%%%%%%%%%%%%%%%%%%%%%%%
\section{Model Management Science} 
How should an analyst use imperfect models to learn from data?\footnote{The challenge of learning from uncertain knowledge is also a fundamental issue in the development of intelligent systems.} What should be the output of such an analysis that can ultimately aid informed decision-making? We address these questions by introducing a general inferential framework for statistical learning and decision-making under uncertainty---which builds on two core ideas: abductive thinking and density-sharpening principle. Some of the defining features of our approach for data analysis, scientific discovery, and decision-making are highlighted below:

% How should an analyst use imperfect models to learn from data? 

% How should an analyst use a misspecified model to make informed decision?

%..contributions ..We conclude by reviewing some key take-away points:

%instinctively guided

~$\bullet$ \textit{Data analysis and science of model management}: No model is perfect, irrespective of how cunningly it is designed. The central problem of statistical model developmental process is to understand how a relatively simple model can evolve into a more complex and mature one in the presence of a new data environment. The principle of density-sharpening assists this model evolution process (thereby helping empirical scientists to abduct): by abductively generating explanations on \textit{why} the presumed model-0 is unfit for the data [playing the role of a quality inspector] and also providing recommendations on \textit{how} to fix the misspecification issues [serving as a policy adviser] in order to make better decisions in new circumstances. 

%and thereby help them to \textit{accelerate} discovery by helping scientists to form novel hypotheses or theory.
\vskip.77em
%Abduction has been largely neglected by empirical scientists.
~$\bullet$  \textit{Discovery and creation of new knowledge}: Abductive data analysts are less interested in \textit{testing} a particular working model. They are mainly interested in conceptual innovation: discovering new pursuitworthy hypotheses based on surprising empirical evidence.\footnote{A largely unexplored topic relative to the vast literature on hypothesis testing. As noted by George E. P. \cite{box2001dis}: ``Much of what we have been doing is adequate for testing but not adequate for discovery.''} The density-sharpening function $d(u;F_0,F)$ picks out `what's new' in the data beyond the current scientific knowledge encoded in $f_0(x)$, thereby helping the scientist to uncover \textit{new unexpected} knowledge from the data using graphical tools. The density-sharpening principle (DSP) provides a learning mechanism that isolates the `known' from the `unknown' and allows us to focus on the newfound pattern in the data, which is the basis for knowledge-creation\footnote{Curious readers are invited to read the paper ``Nobel Turing Challenge: creating the engine for scientific discovery'' by Hiroaki \citeauthor{kitano2021nobel}, where he argued that the single-most-important mission of AI is to accelerate scientific discovery.}.

\vskip.77em
%probability distribution
~$\bullet$  \textit{Abductive inference and decision-making}: The proposed theory of abductive decision-making tackles model uncertainty induced by imprecise, ambiguous, and incomplete knowledge about the underlying probabilistic structure. An abductive-decision support system automatically discovers and explicitly articulates the possible alternatives to the analysts, which forces them to rethink their choices before taking impulsive action; see Supp. note A5.  This style of empirical reasoning and adaptive decision-making could be especially valuable in situations where strategic planners need to take quick action in the face of uncertainty, equipped with approximate subject-matter knowledge.
%investigators 

\section*{Code and Data availability}
All the datasets and R-code written for the analysis are available upon request to the author.
\section*{Ethical Statement}
The Author declares that there is no conflict of interest/competing interests.
\section*{Supplementary material}
For more discussion on how our abductive statistical approach compares to the more traditional Bayesian statistical approach for model misspecification, robustness, and decision-making, see the Supplementary section.

%%%%%%%%%%%%%%%%%%%%%%%%%%%%%%%%%%%%%%%%%%%
\bibliographystyle{Chicago}
\bibliography{ref-bib}

\clearpage

\setcounter{page}{1}
\setcounter{equation}{0}
\renewcommand{\thepage}{S\arabic{page}} 
\renewcommand{\theequation}{E\thesection.\arabic{equation}}
\appendix
\renewcommand{\thesubsection}

\section{Supplementary Notes}
\renewcommand{\theequation}{6.\arabic{equation}}
\appendix
\renewcommand{\thesubsection}{Note \arabic{subsection}}
\vskip2em
This supplementary section contains some additional notes on the connections and differences between the Bayesian statistical approach vs. the Abductive statistical approach to model misspecification, robustness, and decision-making.

% \subsection{The two sides of model-uncertainty: Parametric and nonparametric}
% \label{app:MU2}
\vskip.5em
{\bf A1. The two types of model-uncertainty: Parametric and nonparametric} 
\vskip.3em
It is important to distinguish two main types of model uncertainty:  \textit{parametric} uncertainty and more general \textit{nonparametric shape} uncertainty.
\vskip.25em

~~~1) Parametric uncertainty is a classical scenario in which the model structure is assumed to be known but not the relevant parameter values. In this setup, it is implicitly assumed that the decision-maker is aware of the correct parameterized statistical models $f_0(\cdot;\te)$, which is misspecified in terms of only $\te$.\footnote{As a result, decision theory based on parametric uncertainty is predicated on a highly restrictive model uncertainty assumptions.} We can therefore think of it as a \textit{finite-dimensional} (statistical search) problem, for which we have a number of legacy theories, including Bayesian inference.

\vskip.45em
~~~2) Under nonparametric shape uncertainty, we work with much deeper or more severe uncertainties about the shape of the data-generating model. It is a far more challenging \textit{infinite-dimensional} (statistical search) problem for which there is no established general theory. This paper addresses this concern by offering a `general theory of nonparametric model revision' whose foundation stands on two pillars: the density-sharpening principle and abductive inference. In our theoretical framework, we only have access to a probability model $f_0(x)$ that approximately encodes the decision-maker’s beliefs about the distribution of the observations. Our theory then provides a systematic method for searching a useful class of alternative models $\DS(F_{0},m)$ by `correcting' (or sharpening) the hypothesized model $f_0$ in an automated data-driven manner.

% Thus it is not entirely surprising that the former can also be entirely reformulated using density-sharpening-tools and concepts.

\vskip.5em
{\bf A2.   Bayesian statistical approach vs. Abductive statistical approach}
\vskip.3em
~~~1) Bayesian inference is extremely effective for dealing with parametric model uncertainty. Bayes' law provides a principled method for updating beliefs about a model's parameters.

\textit{Bayes' law}. Given observed data $x=(x_1,\ldots,x_n)$ and the parametrized likelihood function $f_0(x;\te)$, Bayes' multiplicative rule updates belief about $\te$ from the prior to posterior as follows
\beq 
\setlength{\abovedisplayskip}{1em}
\setlength{\belowdisplayskip}{1em}
\label{eq:Bayes}
\wtpi(\te|x) ~ \propto ~ \pi(\te) \prod_{i=1}^n \{ f_0(x_i;\te)\}. \eeq
For more details on standard methods for Bayesian  parametric modeling, see \cite{box1980sampling}.\footnote{For nonparametric Bayesian modeling see \cite{ghosal2017sp}.}

\textit{Bayes decision rule}. Optimal Bayes action is taken by minimizing the expected loss under the posterior: 
\[\hat{a}_{{\rm Bayes}} ~:=~ \argm_{a \in \Abb} \int _\Theta L_a(\te)\, \tilde \pi(\te|x) \dd \te.
\]
For more detailed treatment refer to the standard textbooks like \citet[Chapter 4]{berger2013statistical}.

\vskip.75em

~~~2) Abductive inference, on the other hand, is a powerful mode of statistical reasoning for nonparametric model uncertainty problems. In particular, the proposed density-sharpening law provides a systematic rule for updating the \textit{shape} of a probability density model.

% Bayesian inference provides a principled approach  

% but some prior background knowledge on the parameters. 

% Bayesian decision theory is most powerful when we have some additional background knowledge of the parameters. Bayesian inference learns about the parameter that generated the data.  Bayes rule provides  changing one’s priors
% $\pi(\te_j)$ given new information.

% Bayesian inference for decision making vs. Abductive inference for decision making
% Parameter update based on a subjective prior $\pi(\te)$ in one case and another case focus is more on discovery based on imprecise domain knowledge. 

The abductive decision analysts do not live in a fantasy world  where decision-makers pretend to know the ideal parametrized model for the data. A new class of abductive inference-based decision-theoretic models called ``dyadic models'' are introduced in this paper (see Sec. 2), which allow the analyst to automatically generate a class of probable alternative models from data \textit{without} imposing any prior structural constraints.

\vskip.5em
{\bf A3. Bayes' model synthesis process}
\vskip.3em
The goal of the ``model synthesis problem'' is to answer the following question:

\textit{Given a set of observations, how to systematically go about searching for a model superior to the one the decision-maker initially guessed?} 

\vskip.3em

Bayes model synthesis process (contrast this with the abductive model synthesis process given in Sec. 3.1, method 2) takes into account the uncertainty of $\te$ as follows: 
\vskip.25em

1. Simulate $\te_1,\ldots, \te_B$ from the  posterior distribution $\wtpi(\te|x)$, for some large $B$, say $1000$.

2. Generate a set of plausible parametric models for the data $\{f(x|\te_j)\}_{1\le j\le B}$.

3. Averaging over the posterior distribution: Compute the averaged density  that accounts for the uncertainty of $\te$ (compare this with density-sharpening based bootstrap averaging method,  Eq. \ref{eq:fbar})
\beq  
\bar f_\te(x) = B^{-1}\sum_{j=1}^B f(x|\te_j).
\eeq
which is an approximation to the posterior predictive density
\beq    
\bar f_\te(x) \approx \int_{\Theta} f(x|\te) \, \pi(\te|x) \dd \te
\eeq   
The traditional frequentist point-estimate based $\hat f := f(x;\hat \te)$ underestimates the uncertainty inherent in $\te$, and as a result, it is much `narrower' than the Bayes $\bar f_\te(x)$. By averaging over the posterior distribution, $\bar f_\te(x)$ restores the uncertainty lost when only a single $\hat \te$ is used.

\begin{rem}
Also see Remark 7, where bootstrap is used as the poor man's Bayes posterior probability for each alternative model synthesized from the class $\DS(F_0,m)$.
\end{rem}

% $\bar f(x)$ is in general much wider than the point-estimate-based  frequentist
% $f(x;\hat \te_{{\rm MLE}})$.
% the uncertainty inherent in $\te$
% If we are uncertain about these values, using single-point estimates will underestimate the uncertainty inherent in making these
% predictions, resulting in the spread of the distribution of predictions being too
% narrow.  So, if we average over the posterior distribution, we can restore
% the missing uncertainty. The distribution created by averaging future predictions
% over the posterior densities of all unknown parameters is called the “predictive
% density” in Bayesian analysis.

\vskip.5em
{\bf A4. Awareness of Unawareness}
\vskip.3em
In our abductive model synthesis process, as described in Sections 2 and 3.1, the crucial component is the dyadic model, which is founded on the density sharpening principle. One way to conceptualize dyadic models is as computational agents that are aware of their own \textit{unawareness}. Using this model, decision-makers gain new previously unknown information that had been lurking in the shadows. As the analyst becomes aware of new facts, the belief is nonparametrically updated through the sharpening function $\whd_0(x)$. In other words, the sharpening function alerts decision-makers to their potential ignorance.

% \textit{Significance of dyadic models for decision making.}  

% This is why the ``discovery' of new knowledge from data is so critical in decision-making.

% The proposed decision theory is grounded in a  modern theory of statistical inference, namely the `abductive inference.'

% Virtues of abductive model of decision analysis

% Abductive model building is a continuous iterative process of parameter estimation, uncertainty quantification, and model revision.

% $\bullet$ A general theory of model revision and its application for decision-making under uncertainty. 

\vskip.5em
{\bf A5. Significance of abductive inference for decision making}. 
\vskip.3em
~~~1) \textit{Adaptability}. Reality always carries an element of surprise. 
To make effective decisions in a dynamic uncertain environment, it is critical to ensure that the decision model can withstand surprise; otherwise, it is unfit for use in the real world. The real advantage of using density-sharpening-based dyadic models is that they can recuperate from surprises through automated  structural correction. As a result, an abductive decision rule based on dyadic models can adapt to surprises in the sense that if the true model deviates from the assumed one then still that decision works. This is achieved by  averaging over the plausible  alternative situations suggested by data, as described in section 3.

% In the proposed abductive framework, decisions are made using inferential conclusions drawn from data using density-sharpening methods.

% less vulnerable to surprise. 

% Any subject-matter theory only partially explains reality.  As a result, no decision should be made without first accessing and uncovering the  hidden uncertainties of the theoretical model. This is precisely what abductive inference does for decision-making under uncertainty.
\vskip.4em
~~~2) \textit{Explainability}. Another advantage of abductive inference is that it provides an interpretable and transparent explanation of why and how the real world differs from decision-makers' initial belief about the model, which is of utmost importance when advising on decisions to policymakers.

%%%%%%%%%%%%%%%%%%%%%%%%%%%%%
\vskip.5em
{\bf A6. Bayes, Smooth Bayes, Sharp Bayes, and Robust Bayes} 
\vskip.3em
As an educated guess at the data-generation process, a decision analyst handpicks a class of parametric models $\{f_0(x;\te): \te \in \Theta\}$ with quantile function $Q_0(u;\te)$ and cdf $F_0(x;\te)$.

\begin{defn} [Parametrized sharpening kernel]
\label{def:dpara}
Define the sharpening kernel between the true generating process $f(x)$ and the assumed parametrized $f_0(x;\te)$ as
\beq  
d_\te ~:=~d_\te(u; F, F_0(\cdot;\te))~= ~\dfrac{f(Q_0(u;\te))}{f_0(Q_0(u ; \te);\te)}, ~~\,0<u<1~~~~
\eeq
The corresponding sample estimate is given by
\beq
\wtd_\te ~:=~d_\te(u;\wtF, F_0(\cdot;\te))= \dfrac{\wtf(Q_0(u;\te))}{f_0(Q_0(u ; \te);\te)}, ~~\,0<u<1~~~~\eeq
where $\wtd_\te: \Theta \times [0,1] \rightarrow [0, \infty)$, which connects information in data to parameters of interest. With this definition in hand, we now present an important result.
\end{defn}

{\bf Alternative representation of Bayes Rule}. We express the likelihood-based standard Bayesian posterior update rule for $\te$ as follows
\beq 
\label{eq:sbayes}
\wtpi(\te|x)  ~\propto~  \pi(\te) \exp\big \{ -   \int \wtd_\te \log \wtd_\te \big\}.
\eeq
This is equivalent to Eq. \eqref{eq:Bayes} becuase of the following fact
\bea     
- \int \wtd_\te \log \wtd_\te&=& \int \log\{ f_0(x_i; \te)    \} \dd \wtF~~+~~\text{constant}~~  \\   
&\propto& \sum_{i=1}^n \log \{ f_0(x_i; \te) \}.~~
\eea
The Bayes rule is reformulated in terms of density sharpening kernel $\wtd_\te$ because it offers a  coherent and principled path for generalizing the belief update rule in situations where the probability model (likelihood function) is misspecified.

\vskip.5em

{\bf Smooth Bayes}. Before delving into the Bayesian update rule under model misspecification, we describe ``smooth'' Bayes---an intriguing refinement of traditional Bayes.
Substitute the noisy empirical $\wtd_\te$ with the smoothed $\whd_\te$ (following the method of Sec. 2.2) into Eq. \eqref{eq:sbayes} to get a smoothed version of the Bayes update rule:
\beq 
\label{eq:sbayes2}
\setlength{\abovedisplayskip}{1.25em}
\setlength{\belowdisplayskip}{1.25em}
\whpi(\te|x)  ~\propto~  \pi(\te) \exp\big \{ -   \int \whd_\te \log \whd_\te \big\},\eeq
%\[ \hte_{{\rm MAP}} \,=\, \argmax_\te \pi(\te|x).~~~~~~\]

% The key assumption behind the validity of the above Bayesian rule is that the guessed parametric family $\{f_0(x;\te): \te \in \Theta\}$ \textit{contains} the true data model $f(x)$.  In this case, the Bayesian update can be shown to be the rational way to make a decision using the Savage axioms (Savage, 1954). the posterior $\pi(\te|x)$ guides the optimal action
\textit{Key assumption}. Using the Savage axioms, the Bayesian update can be shown to be the rational way to make a decision when the guessed parametric family $\{f_0(x;\te): \te \in \Theta\}$ \textit{contains} the true data model $f(x)$. However, this is a very stringent requirement that is difficult to meet in practice.  We prefer to operate under more realistic conditions, which allows for $f_0(x;\te)$ to be misspecified.

\vskip.5em
{\bf Sharp Bayes}.  What if the analysts' a priori chosen family $\{f_0( \cdot;\te): \te \in \Theta\}$ \textit{does not} contain the actual data generating model $f(x)$? It is well known that Bayes' update exhibits undesirable characteristics under model misspecification. 

\begin{defn} [Generalized $d$-posteriors]
\label{def:dpara}
Define the following divergence-based generalized posterior density function
\beq 
\label{eq:dpost}
\pi(\te|x)  ~\propto~  \pi(\te) \exp\big \{ -  I_{\psi}(F,F_0(\cdot; \te))\, \big\}\eeq
where $I_{\psi}(F,F_0(\cdot; \te)) $ is the Csisz{\'a}r class of divergence measure between the true data generator $f$ and the assumed misspecified class $f_0( \cdot ; \te)$. We refer to \eqref{eq:dpost} as $d$-posterior, because we can rewrite it using Eq. (16) as
\beq 
\pi(\te|x) ~\propto~ \pi(\te) \exp\big \{ - \int \psi \circ d_\te \big\}. ~~\eeq
\end{defn}
whose sample estimate is given by
\beq 
\label{eq:czpost}
\wtpi(\te|x) ~\propto~ \pi(\te) \exp\big \{ - \int \psi \circ \wtd_\te \big\}. ~~\eeq

In a remarkable result, \cite{bissiri2016general}\footnote{After some algebraic manipulation,  it is not difficult to show that  the main result of Bissiri et al. is equivalent to \eqref{eq:czpost}.} showed that 
\eqref{eq:czpost} provides a valid coherent rule for revising prior beliefs about the parameters of a model that is misspecified.  Sharp-Bayes is the name given to this density-sharpening-based generalized Bayes update rule.

\begin{rem}[The key idea]
The information in the observed data $x_1,\ldots,x_n$ is connected with the parameter of interest $\te$ via functionals of the sharpening function $\wtd_\te$, instead of the conventional likelihood function, whose precise probability form is never known in practice. 
\end{rem}

 %Robust Bayes under model misspecification. 
{\bf Robust Bayes}. 
For outlier-resistant robust Bayesian analysis choose total variation divergence, a special case of Csisz{\'a}r class with $\psi(x)=|x-1|$ in \eqref{eq:czpost} 
\beq 
\label{eq:czpostr1}
\wtpi(\te|x) ~\propto~ \pi(\te) \exp\big \{ - \textstyle \int | \wtd_\te - 1 | \big\}. \eeq
Another particularly useful class of measures for robust Bayesian analysis is R{\'e}nyi $\al$-divergence, defined as
\beq
\label{eq:reyni}
R_{\al}(F,F_0)\,=\, \dfrac{1}{\al (1-\al)}\textstyle \Big( 1 - \int  ( f/f_0 )^{\al} \dd F_0 \Big), ~~~\al \in \cR \char`\\ \{0,1\}.
\eeq
%susceptible to outlying observations
It is a robust discrepancy measure between $f$ and the imperfect $f_0$ whose nonparametric estimation can be done by expressing it as a functional of the sharpening function:
\[R_{\al}(\wtF, F_0(\cdot;\te))\,=\,\dfrac{1}{\al (1-\al)}\textstyle \Big( 1 - \int  \wtd_\te^\al \Big), ~~~\al \in \cR \char`\\ \{0,1\}.\]
The following is the associated posterior belief update rule:
\beq  \wtpi(\te|x) ~\propto~ \pi(\te) \exp\Big \{  -  \frac{1}{\al (1-\al)} ( 1 - \textstyle{\int }\,  \wtd_\te^\al ) \Big\} \eeq
The value of $\al \in [0.50, 0.75]$ is commonly used to provide good robustness protection against outliers without losing too much efficiency.

\vskip.5em
{\bf Two major conclusions}: 

(1) Knowing the `gap' between the sample distribution and the true data generator (as captured by $\wtd_\te$) is sufficient to produce posterior beliefs, obviating the need to know the exact probabilistic form of the true likelihood function,  which decision-makers almost never know in real-world scenarios.

(2) The fundamental object of statistical inference is not the guessed misspecified parametric model $f_0( \cdot;\te)$ nor the unknown $f$, but the `gap' between them, $d_\te$.

%%%%%%%%%%%%%%%%%%%%%%%%%%%%%%%%%%%%%%%
\vskip.5em
{\bf A7. Addressing Prior misspecification}  
\vskip.3em
The information-theoretic generalized Bayes rule, presented in the previous note, is still not fully satisfactory because it is rigidly based on assumed subjective prior $\pi(\te)$.\footnote{For a more detailed account of  ``subjective'' probability theory see the classic book by \cite{de74theorySP} and also \cite{lad1996SP}.}   Thus it is critical to investigate the robustness of statistical decisions in a reasonable neighborhood around the presumed prior, which can be operationalized 
through the density sharpening principle; see, for example, \cite{deep18naturesupp}.  Instead of making critical decisions based solely on analysts' vague subjective specifications, this allows for prior misspecification.

\vskip.4em
Notes A6 and A7 showcase how concept density-sharpening principles can unify both the classical and the most advanced versions of Bayesian inference using common terminology and notation -- a novel contribution in and of itself.

\end{document}